\documentclass{article}

\usepackage{arxiv}

\usepackage[utf8]{inputenc} % allow utf-8 input
\usepackage[T1]{fontenc}    % use 8-bit T1 fonts
\usepackage{hyperref}       % hyperlinks
\usepackage{url}            % simple URL typesetting
\usepackage{booktabs}       % professional-quality tables
\usepackage{amsfonts}       % blackboard math symbols
\usepackage{nicefrac}       % compact symbols for 1/2, etc.
\usepackage{microtype}      % microtypography
\usepackage{lipsum}		% Can be removed after putting your text content
\usepackage{graphicx}
\usepackage{natbib}
\usepackage{doi}

\title{\Large Fair Models in Credit: Intersectional Discrimination and the Amplification of Inequity}

\author{ 
    {\hspace{1mm}Savina D. Kim} \thanks{S.Kim is affiliated with the University of Edinburgh Business School and the Centre for Technomoral Futures at the Edinburgh Futures Institute, United Kingdom} \\
	University of Edinburgh\\
	Edinburgh, United Kingdom \\
 	\And
    {\hspace{1mm}Stefan Lessmann} \\
	Humboldt University of Berlin\\
	Berlin, Germany \\
	\And
    {\hspace{1mm}Galina Andreeva} \\
	University of Edinburgh\\
	Edinburgh, United Kingdom \\
	\And
    {\hspace{1mm}Michael Rovatsos} \\
	University of Edinburgh\\
	Edinburgh, United Kingdom \\
}

\renewcommand{\shorttitle}{}

\hypersetup{
pdftitle={Fair Models in Credit: Intersectional Discrimination and the Amplification of Inequity},
pdfsubject={stat.ML},
pdfauthor={Savina D.~Kim, Stefan Lessmann, Galina Andreeva, Michael Rovatsos},
pdfkeywords={Fairness, Discrimination, Intersectionality, Credit scoring, FinTech, Open Banking, Alternative credit data},
}

\begin{document}
\maketitle

\begin{abstract}
The increasing usage of new data sources and machine learning (ML) technology in credit modeling raises concerns with regards to potentially unfair decision-making that rely on protected characteristics (e.g., race, sex, age) or other socio-economic and demographic data. The authors demonstrate the impact of such algorithmic bias in the microfinance context. Difficulties in assessing credit are disproportionately experienced among vulnerable groups, however, very little is known about inequities in credit allocation between groups defined, not only by single, but by multiple and intersecting social categories. Drawing from the intersectionality paradigm, the study examines intersectional horizontal inequities in credit access by gender, age, marital status, single parent status and number of children. This paper utilizes data from the Spanish microfinance market as its context to demonstrate how pluralistic realities and intersectional identities can shape patterns of credit allocation when using automated decision-making systems. With ML technology being oblivious to societal good or bad, we find that a more thorough examination of intersectionality can enhance the algorithmic fairness lens to more authentically empower action for equitable outcomes and present a fairer path forward. We demonstrate that while on a high-level, fairness may exist superficially, unfairness can exacerbate at lower levels given combinatorial effects; in other words, the core fairness problem may be more complicated than current literature demonstrates. We find that in addition to legally protected characteristics, sensitive attributes such as single parent status and number of children can result in imbalanced harm. We discuss the implications of these findings for the financial services industry. \end{abstract}

% keywords can be removed

\section{Introduction}

Concern that technological advancements can have unintended discriminatory consequences is a persistent theme found in the history of computation and data analytics. This speaks to the threat that big data, algorithmic processes and automated decision-making (ADM) systems when applied to certain problems, have the power to disguise and exacerbate unjust distribution of crucial liberal goods such as rights, opportunities and wealth. This can include whether an individual qualifies for credit or employment opportunities, are released on bail, or which housing advertisements they are shown \citep{Smith2016}. 

Research on consequential bias and unfairness in computing is driven, in part, by the rapid propagation of devices and online platforms which facilitate the collection of vast amounts of social and behavioral data \citep{Boyd2012}. One area which has seen increased investment into this space is financial services, particularly by financial technology (FinTech) companies looking to provide alternatives to traditional banking services. Leveraging vast digital platforms and armored with a myriad of new data types, FinTechs can capture in-depth, personalized insights into prospective borrowers thus providing them a competitive edge. This is especially the case in riskier borrower segments and for smaller loan sizes where screening using automated credit scoring technologies is more profitable \citep{Einav2013} alongside having a more favorable cost structure by operating often exclusively in an online environment (e.g., no costly branch network, no capital or liquidity requirements). Through improved credit risk management, FinTechs have been able to expand their pool of prospective borrowers. This is particularly prominent with the unbanked population or those with “thin-files” who lack credit histories to be evaluated on \citep{Agarwal2019, Brevoort2016} and who typically have had to seek finance at high-cost lending institutions such as payday loan companies \citep{Bhutta2015}. This marketplace of alternative credit providers, providing small loans with almost instantaneous application processes and sometimes without the need for a credit score, serves as a viable alternative to mainstream banking particularly for those who may not satisfy a bank’s traditional lending requirements. 

However, in light of the recent growth of FinTech creditors, there has been heightened interest among the regulatory community and academic researchers who are wary of the novel risks that come with these technological and data-driven advancements. To date, assessment of machine learning (ML) processes has revealed the potential for unjust (and illegal) discriminatory outcomes which pose significant risks for the broader public and particularly those already systematically disadvantaged (i.e., vulnerable sociodemographic groups) which services claim to benefit the most. This view is also reflected in the current regulatory discussion, which calls for greater consumer protection initiatives with regards to ‘responsible’ and ‘ethical’ AI usage \citep{Hacker2018, Langenbucher2020, Purificato2022, Wachter2021, Heikkilaarchive2022}. To this end, the growing usage of granular, more intrusive data alongside complex, state-of-the-art modelling processes raises major concerns regarding the viability of online lending platform businesses for the greater good and thus merit attention.

The objective of our study is to explore the issue of fairness in the credit context of microfinance on a real-world dataset. We investigate the discriminatory potential of alternative credit services using data on over a hundred thousand unsecured consumer loans sourced from a microloan lending platform servicing the Spanish market. The study looks to provide both empirical evidence and a sociological discussion on the performance of automated credit models using alternative credit data and its implications. Thus, it is not only a specific investigation of AI-based credit scoring but also a general exploration of the promises and pitfalls of AI and big data, particularly the ‘all data is credit data’ approach. We focus on the concept of financial inclusion (i.e. whether people can have convenient and affordable access to capital) where credit scoring influences both its magnitude and quality. Achieving an ideal level of financial inclusion in this digital era remains a demanding task, particularly in areas of the marketplace where the credit system is not well established and regulated (i.e., microcredit) and information on sensitive attributes are lacking. 

Our analysis is unique in comparison to previous studies. Firstly, we incorporate both single parent status and number of children as sensitive attributes, rarely seen in fair ML works. This is assessed alongside gender, age and marital status in the analysis of difficulties accessing microlending services. We highlight the need to expand the characteristics considered in such analyses, beyond only those which are legally protected, owing to the ability of increasingly complex ML models to uncover hidden patterns in large data with high predictive accuracy. Secondly, while it is common for quantitative research to assess the independent impact of legally protected attributes (e.g., age, gender, race or ethnicity) on lending decisions, the literature suggests that this approach overlooks the complexity of intersectionality and potential amplification of harm it poses. The importance of this approach is highlighted further by intersectionality theory, which examines how various socially and culturally constructed classifications do not act individually but rather interplay on multiple levels, thereby creating a system of oppression that contributes to societal inequalities. We term this concept “multi-multi,” representing the use of multiple and multinomial combinations of SAs to assess fairness in-depth. We contribute to the literature showing that unfairness can worsen as multiple SAs are combined and deeper levels are considered; in other words, the core problem is more complicated than what the current literature demonstrates. Thirdly, we present a novel dataset which encompasses a diverse range of features sourced from multiple third-party data providers, including customer-provided personal and sociodemographic information, bank transaction features (e.g., features engineered based on income, loan-to-income, transactional text), data from the credit bureau agency, user agent details (e.g., browser used, operating system) and digital footprint data based on the device used during application process (e.g., language, screen resolution). 

In the following, we first provide a review of the literature on alternative credit data usage, ML and algorithmic fairness in consumer credit modelling (Section \ref{sec:Related Work}), though the latter is taken mainly from computer science literature with a predominately application-agnostic perspective. In Section \ref{sec:Methods}, we introduce the methodology and data employed in the analysis. The predictive models are segmented first by data inclusion with preliminary analysis using interpretability methods and the results of the fairness evaluation are discussed further in the analysis (Section \ref{sec:Results}). Following the final conclusions, we also highlight the potential implications for alternative credit lenders, its customers and regulators (Section \ref{sec:Discussion}). 

\section{Related Work}
\label{sec:Related Work}

\subsection{Credit Scoring with FinTechs}

Traditional credit scoring models typically include sociodemographic data, which includes information on the applicant’s financial situation, employment status and education level. Alternative lenders provide access to credit for borrowers outside of traditional banking systems. With greater flexibility in designing their own credit scoring models and aided by less legal scrutiny, these models can leverage all kinds of available ‘non-traditional’ digital information to infer a prospective borrower’s creditworthiness \citep{Aitken2017}, including but not limited to: shopping history, utility bills, mobile usage, and even social media. 

One major source of data that has not yet been used to the full extent is massive fine-grained payment data \citep{Tobback2019}. Bank card transaction records are playing an increasingly important role in credit scoring models given their ability to capture an individual’s actions and/or interactions to form a behavioral profile \citep{Shmueli2016, Thompson2020, vissing2012}. This is not only beneficial but also easily accessible given banks have this information at their disposal if the applicant is an existing (credit) client. Even without being a direct customer, the new European Payment Services Directive or Open Banking in the UK which came into effect in 2018 has encouraged greater mobility of consumers’ payment data through third-party providers which can be authorized at the customer’s request. Studies have shown the efficacy of using payment data to supplement (or even replace) data from other data sources such as credit bureaus \citep{Omarini2018, Remolina2019}. For example, \citet{Foos2010} examined the influence of credit line usage and checking account balance on the default risk of borrowers. The authors found that features derived from account activity significantly enhanced default predictions. \citet{Khandani2010} analyzed patterns in consumer expenditures, savings and debt payments to predict credit card delinquencies. Similarly, \citet{Bellotti2013} predicted credit defaults using dynamic models with monthly account behavioral records. Even grocery shopping data has been shown to be informative when predicting credit card repayment \citep{Lee2022}. Other types of alternative credit data such as digital footprints (i.e., web browsing activity) was captured in \citet{Berg2020} for predicting default, showing that digital data can complement credit bureau information for superior lending decisions. This is similarly shown with mobile usage and communications data, such as the number and duration of calls \citep{Pedro2015, bjorkegren2020} and even social data, such as the number of friends, social ties, and social media disclosure, which are found to correlate with default risk \citep{Lin2012, Zhang2016}.

In addition to novel and diverse data sources, advancements in ML have also enabled more sophisticated credit models \citep{Baesens2003, Fuster2019}. This has dramatically reshaped traditional practices for screening borrowers by enabling models to incorporate the aforementioned, seemingly irrelevant borrower characteristics (i.e., non-standard features) to probe into the complex relationships between creditworthiness and the borrower’s profile \citep{Lessmann2015}. \citet{Morse2015}'s review of the literature developing around FinTech lending with a focus on the type of technologies employed has shown its ability to mitigate information frictions in lending. The author suggests that improved capturing of soft information contained in proximity information and thus improved profiling of applicants can expand access to or pricing of credit. Credit scoring research has often focused on modelling techniques (and only more recently to a greater extent on data input), highlighting the ability of AI-based credit scoring models to significantly outperform the industry standard logistic regression with regards to predictive power. This includes ML such as neural networks, classification trees, random forests and support vector machines to name a few \citet{vieira2019, Khandani2010, Kruppa2013, Malagon2022} and related ensembles classifiers \citep{Lessmann2015}.

\subsection{Fair Lending}

While many researchers and practitioners express optimism about these data-fortified, state-of-the-art models and its promise of new opportunities for the field, others express apprehension about their risks. They call for caution, particularly with regards to how algorithmic systems may stand to undermine fairness when applied in real world settings and their potential to discriminate. As one White House report puts it, there is an acute risk of ‘unintentional perpetuation and promotion of historical biases,’ especially in cases ‘where a feedback loop causes bias in inputs or results of the past to replicate itself in the outputs of an algorithmic system’ \citep[p. 8]{Smith2016}. Needless to say, there is no shortage of historical accounts of lending discrimination \citep{Agier2013_mfandgender, Bayer2018, Berkovec1998, Black1978, Bocian2008,Chen2017, Chen2020, Cheng2015, Courchane2007, Munnell1996, Pope2011}. 

Especially when implemented with large datasets, ML algorithms are likely to better capture the structural relationship between observable characteristics and default \citep{Jansen2023}. However, the consequential effect on discrimination can go one of two ways. On one hand, combining advanced modeling techniques with non-standard data permits the inclusion of small borrowers traditionally overlooked when using standard screening techniques \citep{Berg2020}. On the other hand, the model and empirical evidence in \citet{Fuster2022}, for example, highlights that ML can increase rate disparity across different groups of borrowers, benefiting White and Asian borrowers disproportionately relative to Black and Hispanic borrowers. Compared to standard parametric scoring models, such as logistic regression, ML models introduce an additional flexibility which is useful for improving out-of-sample classification accuracy. However, these gains are often not homogeneously distributed across applicants such that minorities may be affected by a triangulation effect \citep{Hurlin2022}. This occurs when non-linear associations between the features proxy for sensitive characteristics, thereby “de-anonymizing” identities using non-sensitive information.

In response, researchers have looked to strategies for combatting the negative impacts of algorithmic decision-making, led majorly by computer scientists and sociologists focused on the mathematical foundations and societal impacts of ML, respectively. A wide range of computational solutions to problems of bias and unfairness have been proposed \citep{Berk2017, Calders2010, Dwork2012, Feldman2015, Hajian2012, Kamiran2012}; including the advancement of statistical definitions of fairness for operationalization within computation systems and debating which best serve the fairness task \citep{Friedler2021, Kilbertus2017, Kleinberg2016}. \citet{Narayanan2018} examines over 21 fairness frameworks, highlighting the values and politics weighed within the definitions and the substantial moral assumptions implied by its mathematical formulas. We refer the reader to \citet{Majumder2021} and \citet{Verma2018} for a more comprehensive overview. 

However, a major obstacle in this field is the mathematical impossibility of fulfilling all definitions of fairness simultaneously, particularly with different base rates across subgroups. This is formally known as the impossibility theorem \citep{Friedler2021}, and consequently poses the dilemma of choosing which fairness metric to use. Having explored different fairness trade-offs \citep{Kleinberg2016}, the inevitable conclusion is that data scientists must decide which metric to prioritize in a given model based on the desired gain (or loss) in performance. In this predicament, multiple authors highlight the cruciality of social context when assessing appropriate understandings and implementations of fairness solutions \citep{Ferrer2021, Lepri2018, Wong2020}. Ultimately, deciding which fairness definition and consequently which metric is most appropriate depends on the task at hand or domain in which the model operates. For example, a model which supports decision-making that may limit an individual's life opportunities (e.g., default or recidivism prediction) may want to prioritize the minimization of false positive rates for discriminatory groups, whereas a model which implies a favorable outcome may desire minimizing false negative rates \citep{Bellamy2019}. We look to provide this in the context of microfinance lending. 

\subsection{Intersectionality}

Lastly, we highlight a major limitation to anti-discrimination discourse which is the tendency towards single-axis thinking. This insight is central to the notion of intersectionality pioneered by Kimberlé Crenshaw in her foundational work \citep{Crenshaw1989}. Intersectionality is a lens for examining societal unfairness which considers that an individual’s identity can hold multiple dimensions which do not exist in isolation. Rather they can interact collectively in a way which affects individuals’ experiences and behaviors in relation to inequality, injustice, exploitation and oppression \citep{Gutterman2022}. Crenshaw highlights the law’s propensity to focus on only one axis of discrimination at a time, thus potentially resulting in harm that is greater than the sum of the parts for those whose experiences span multiple dimensions. The example provided by Crenshaw are the shortcomings of Black women in employment discrimination cases, who are often unsuccessful largely due to the fact that courts compare their claims against the experiences faced by similarly situated Black men (with regards to racial discrimination cases) or White women (with regards to gender discrimination cases). However, these seemingly analogous groups enjoy systematic advantages along at least one historically contingent dimension: gender for Black men and race for White women, therefore does not adequately account for being multiply-oppressed \citep{Hoffmann2019fairnessfails}.

Like the courts, empirical work on data and algorithmic discrimination have yet to fully embrace the lessons learned from intersectionality research originating in the social sciences, where it serves as the language needed for “examining interconnections and interdependencies between social categories and systems \citep{Atewologun2018}”. While efforts to isolate and mitigate unfairness in ADM systems are becoming common, fair ML work within the limits of financial services applicability tends to focus on a limited set of pre-defined protected characteristics (e.g., race, gender) and often one per dataset assessed \citep{Kearns2018}. For example, \citet{Das2021} presents a case study of a fairness-aware ML system in finance by showing how to apply bias measurement and mitigation at different stages in the ML pipeline, namely, pre-training and post-training; however only uses the single feature of gender. \citet{Hurlin2022} similarly illustrates a fairness assessment framework by testing for gender discrimination in a database of retail borrowers using the German Credit dataset. \citet{Hu2000} examines human evaluators’ decision-making using real-world data of an online microlending platform. By developing a structural econometric model to capture decision dynamics, the authors find two types of biases in gender, i.e. preference-based and belief-based bias. \citet{Stevens2020} evaluate four bias mitigation algorithms (learning fair representations, reweighing, Equality of Odds, and Reject Option based Classification) based on a standard XGBoost classifier to build an explainable and fair prediction model using a real-world loan dataset sourced from Kiva.org. They too only focus on gender. \citet{Kozodoi2022} provides the most comprehensive benchmarking analysis by empirically comparing a range of different fairness processors along several performance criteria using seven real-world credit scoring data sets. The protected characteristic used is a binary indicator for age, with 25 as the threshold dividing young versus aged groups, reflecting the work of \citet{Kamiran2009}.

While the majority of work in this space focuses on single binary demographic attributes, there is a limited few which consider multiple attributes simultaneously. For example, \citet{Teodorescu2021} employs the popular credit decision dataset UCI Bank Marketing Dataset and applies several fairness criteria to show that many typical classification algorithms cannot satisfy more than one fairness criterion at a time when considering more than one attribute, using marital status and age. \citet{Singh2022} creates a bias mitigation strategy coined DualFair targeted at the mortgage domain using the HMDA dataset, which debiases using data oversampling and undersampling techniques. To extend their approach to intersectional fairness, the authors subdivide datasets using three protected characteristics (e.g., race, sex, and ethnicity). Through the combination of these sensitive parameters, a total of 27 unique datasets (i.e., groups) are formed and utilized throughout the pipeline. \citet{Kearns2018} propose methods for combining protected characteristics and certifying fairness across multiple subgroups to prevent “fairness gerrymandering,” which our work similarly reflects. From a humanities perspective, \citet{Noble2018} critiques the behavior of the Google search engine by examining the search results for terms relating to women, people of color, and their intersections, e.g., “Black girls” using an intersectional lens. \citet{Buolamwini2018} identify another issue within facial recognition systems, exemplifying that the problematic distinctions between groups are not only limited to pre-existing or pre-targeted categories but can also be fashioned through interactions between labels within the system, stating ‘increasing phenotypic and demographic representation in face datasets and algorithmic evaluation (p. 12).’ This alludes to the concern that attributes outside the limitations of those legally protected can have disadvantageous effects on individuals at each intersection of the affected dimensions, from gender, race, nationality, sexual orientation, disability status, socioeconomic class to numerous other facets of social stratification. We refer to these going forward as ‘sensitive attributes’ (SA). To determine these additional categories, we must first ask ourselves what are the identities that overlap or intersect to make people vulnerable to the rights violation (i.e., fair access to credit) being discussed; for example, a poor Black female, single mother, rural Latina or trans-women of color \citep{Gutterman2022}.

\section{Data and Methods}
\label{sec:Methods}

\subsection{Dataset Description}

We perform our experiment on a dataset sourced between 2013 to 2019 from a FinTech microloan provider, which has previously never been considered in the fairness-aware ML literature. The lender provides microloans up to 700€ in the Spanish market. The dataset contains 129,457 instances (i.e., borrowers) and 3,123 engineered features describing the borrower, along with a binary classification of each individual as having repaid or defaulted on their loan indicating good or bad credit risk. The SAs included in the original dataset are gender as female or male, age as an integer, personal status as single, married, divorced, or widowed, single parent status as a binary indicator and number of children as an integer. For our analysis, age is discretized into categories ‘aged’ (at least 25 years old) and ‘young’ based on the study by \citet{Kamiran2009} which showed this discretization provided for the most discriminatory possibilities. Number of children is discretized into three categories which include ‘no children’, ‘1-2 children’ or ‘3+ children’. And customers who are ‘widowed’ are removed from the analysis due to the small sample size (<2\% of the whole population). Please see Table \ref{table:table1} for descriptive statistics, which displays the proportions of borrowers based on their demographic characteristics and their relative loan performances. The descriptive statistics regarding the intersectionality combinations can be found in the Appendix (Table \ref{table:apx1}, \ref{table:apx2}).

We isolate these sociodemographic characteristics noting the empirical work in other domains which have shown their importance. For example, ‘parents’ are a heterogeneous group spanning a range of chronological ages and socio-economic contexts and include members who are simultaneously exposed to challenges faced by other vulnerable groups on the basis of gender (i.e., female parent), age (i.e., young parent) or marital status (i.e., single parent). Consider the situation of single parent households, who are at high risk of financial hardship \citep{Stack2018}, who tend to be overwhelmingly women \citep{Andersen2023} and generally are members of several other familiar categories including women, immigrants \citep{McLanahan1994}, low socioeconomic class and workers that carry their own unique risks of vulnerability. They also tend to suffer potential rights abuses such as lack of benefits, lower wages \citep{Reed2005} and economic marginalization; for example, they face significant discrimination relative to heterosexual couples in rental markets \citep{Lauster2011, Murchie2018} and labor markets \citep{Dowd1995}. Members of this group must deal with additional pressures beyond those mentioned above due to their single parent status, including the devaluation of unpaid care, limited employment opportunities and lack of flexible and affordable childcare services, which have an overall impact on their financial health and status \citep{McLanahan1994}. Furthermore, number of children has also been shown to be an impactful factor of financial stress \citep{Eling2021, Lundberg1994, Worthington2006}, supporting our claim that these additional attributes should be taken into account when considering ethical reforms of ADM systems in financial services. 

\begin{table}[h]
\includegraphics[width=9cm]{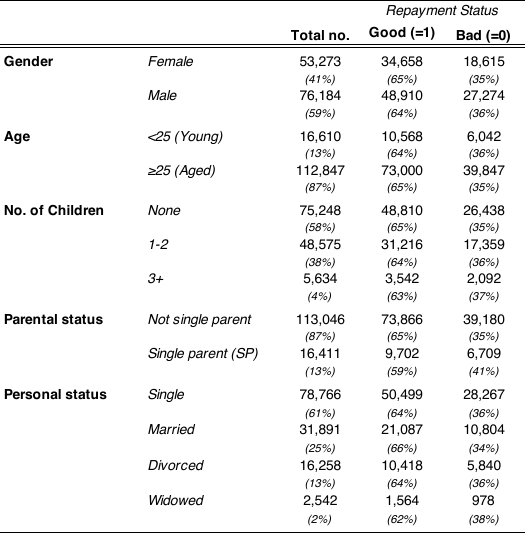}
\centering
\caption{Description of borrowers’ demographic characteristics and repayment status.}
\label{table:table1}
\end{table}

We make a few initial observations. First, we observe that males, aged individuals, those with no children, not single parents, and those of single personal status make up most of the customer base. However, among the different demographic groups, the relative repayment status is similar (approximately 65\%), with the exception of the single parent group where those who are not single parents have a repayment status of 65\% and those who are single parents of 59\%. However, greater disparities in relative repayment are seen along intersectional groupings. The lowest repayment groups are those who are young and single parents (52\%), young and married (52\%), young with 1-2 children (52\%), and single parents with children (58-59\%). The highest repayment groups are married individuals with 1-2 children (68\%), married females (67\%), not single parents and female (66\%), no children and female (66\%) and aged and married (66\%). We can already see in the original data the diverging repayment behaviors among those who are young and have children, particularly single parents, which becomes relevant in the later analyses.

The independent features give information about the loan requests, the borrowers, and data sourced from third party data providers (i.e., credit bureau, bank transaction features, digital footprints). Table \ref{table:table2} shows a list of the data types, categories, number of features per category and a short description. The data types become relevant when administrating different variations of the model to see impact on fairness based on data exclusion. In order to maintain the diversity of data types sourced across various data partners, all major categories are preserved during the preprocessing steps which are detailed next. 

\begin{table}[h]
\includegraphics[width=15cm]{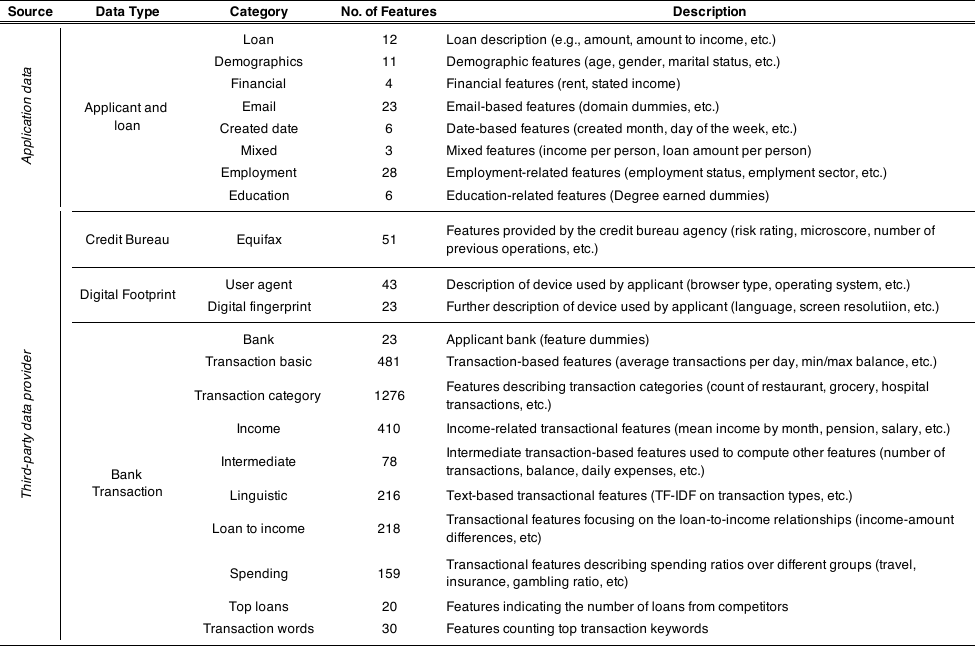}
\centering
\caption{Original input features.}
\label{table:table2}
\end{table}

\subsection{Prediction System}

The methodology applied in this study consists of five major parts, which looks to shed light on the application of discrimination towards underserved minority groups. A standard ML workflow is utilized, including data collection, data preparation, feature selection, explainability and modern training; further details on each of these steps are provided in the following. In the first step, pre-processing, we transform and filter several original features to enhance the analysis. We first modify the input data according to any data-specific needs: removing features that should not be used for classification (e.g., SAs), removing or imputing any missing data, and adding derived features. To enable an analysis of fairness based on multiple SAs simultaneously, we add combined categories (e.g., attribute “age-gender” with values like “Young-Female”) to make up the processed dataset. For binary SAs, we modify the given privileged group to be represented as one and the unprivileged group as zero. These are designated arbitrarily but are motivated by both empirical and real world events highlighting cases where certain groups are given an undue advantage; for example, females are given higher mean interest rates, lower loan amounts, and lower property values compared to those of males \citep{Singh2022} and Apple Pay was recently publicly criticized for setting credit limits for female users at a much lower level than for otherwise comparable male users \citep{Vigdor2019}. Note that bias can also be bi-directional, e.g., bias in either direction is undesirable such that we do not want a gender imbalance, irrespective of gender.  

In the second step, we look to reduce the high-dimensional dataset which holds over three thousand features into a more manageable size. First, a correlation filter is applied where features with greater than 0.7 correlation value are removed. Next, the list of relevant features is further reduced while maintaining the diversity of data types found in the original dataset. For the feature removal step, categories with over 30 features are used to construct a random forest model (number of estimators = 100). The feature importance values for predictive accuracy are extracted with the top 30 features retained; for subgroups with less than 30 features, all features are retained. After feature reduction, the processed dataset holds a reduced number of 376 features while maintaining all data categories found in the original source. 

In the third step, we construct the baseline model based on XGBoost. First, hyperparameter tuning is performed using Bayes Optimizer with 30 iterations to acquire the best parameters. Next, the training set is created by randomly sampling 70\% of the data, with the remaining 30\% kept for testing to assess performance. Downsampling is used to overcome the class imbalance problem to a ratio of one-to-one and afterwards, stratified 5-fold cross-validation is applied to shuffle and split the balanced sample into five folds while preserving the percentage of samples for each class. For each fold, the test set of 38,837 loans is evaluated based on accuracy, precision, recall, and the Area Under the Curve (AUC). By averaging these evaluation metrics across the 5 folds, we provide an accurate performance overview of the baseline model. 

In the fourth step, an interpretability analysis of the model obtained from step three is conducted for the purposes of feature selection and importance. For the baseline model, interpretability is assessed using feature importance plots based on the Shapley method \citep{Lundberg2017}. To expand on alternative credit data impact, we create three additional variations of the model on the basis of data input: (1) ‘traditional’ which includes only customer and loan information with credit bureau input which are used in traditional credit scoring models (2) ‘traditional \& bank transaction’ which includes data used in the traditional model in addition to data types within the bank transaction category, (3) ‘traditional \& digital footprint’ which includes data used in the traditional model in addition to data types within the digital footprint category. As a result, we conclude by producing four variations of data-varied models, including our baseline model (i.e., all data types).

\begin{figure}[h]
\includegraphics[width=\linewidth]{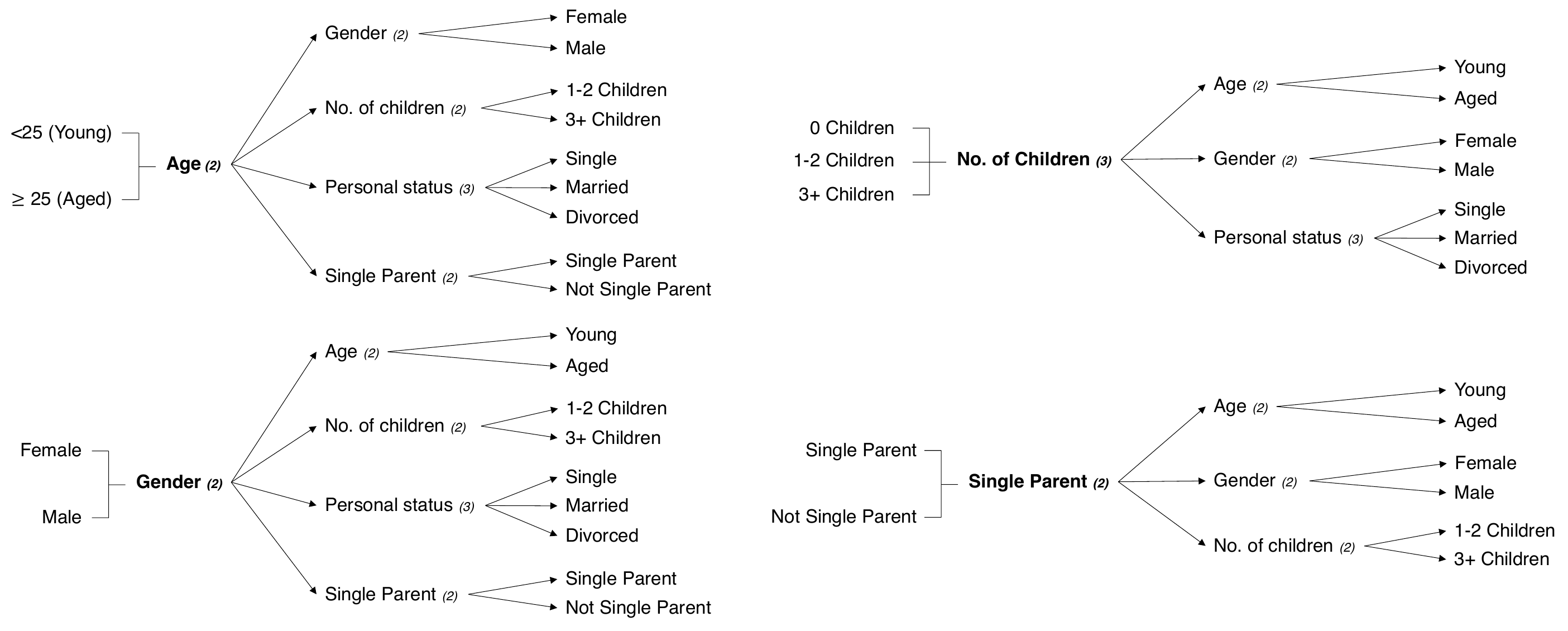}
\centering
\caption{The range of two-part combinations of sensitive attributes considered in the analysis; includes multiple, multinomial characteristics (i.e., ‘multi-multi’). }
\label{fig:fig1}
\end{figure}

In the final step, an investigation into model fairness is conducted. The four variations are assessed to understand the effect of data granularity and non-financial information on deemed ‘higher-level’ fairness outcomes. As such, the effect of the data types used is observed by looking at the confusion matrices of each of the five SAs individually; specifically, the fairness of each model is assessed based on the conditions defined in Table \ref{table:table3}. Finally, we use the baseline model to determine the impact of intersectionality, or what we coin as the ‘lower-level’ fairness outcomes in relation to two-part SA combinations (Figure \ref{fig:fig1}). Given that the fairness metrics vary considerably for different test sets, a bootstrapping procedure (using 500 iterations) is employed to take this variability into account when analyzing the results. Therefore, the values of the fairness metrics are presented as the average value across all iterations.

\subsection{Fairness Metrics}

Existing group fairness metrics can be divided into three categories: independence, separation, and sufficiency \citep{Mehrabi2021}. According to \citet{Friedler2021}, this can be viewed in relation to two different world views: "We are all equal" versus "What you see is what you get." On one hand, we can consider all groups to be equal such that differences found in the data are a consequence of a structural bias that impacts the data acquisition. On the other hand, we can assume that real differences in behavior do in fact exist across groups and they can be captured even without bias. The first world view, where all groups are inherently equal, implies that positive rates (i.e., probability of being assigned the favorable outcome) should be equal across all groups; therefore, the predicted outcome is independent from the SA, also termed independence. The second worldview, where behavioral differences do exist across groups, implies a model is only fair when the classification accuracies (i.e., error rates) are equal across the groups; this is covered by fairness classes separation and sufficiency. 

To be as clear as possible, we provide the following notation. We refer to \emph{A} as the categorical random variable representing the SA. We label with X all other (non-sensitive) random variables that describe the borrower which the algorithm considers to reach an approve or reject decision $\hat{Y}=f(X,A)\in{0,1}$ and label $Y\in{0,1}$ refers to the ‘actual’ or ground truth target variable predicted. This is typically done by minimizing a loss function represented as $\mathcal{L}(Y,\hat{Y})$. $\widetilde{X}\ =\ ({\widetilde{X}}_1,...,{\widetilde{X}}_d)\ =\ (X,A)$ collectively represent all independent variables involved in the problem. Lowercase letters are used to denote specific realizations of random variables, such that $\left\{(x_1,\ y_1),...,(x_n,\ y_n)\right\}$ represents a dataset of n independent realizations of \emph{(X,Y)}. Calligraphic symbols are used to refer to domain spaces, e.g., $\mathcal{X}$ denotes the space where features X live. 

\textbf{Independence} aims to ensure the outcomes of an algorithm are independent of the groups that the individuals belong to \citep{Dwork2011, Feldman2015}, such that:

\newcommand\independent{\protect\mathpalette{\protect\independenT}{\perp}}
\def\independenT#1#2{\mathrel{\rlap{$#1#2$}\mkern2mu{#1#2}}}
\begin{equation}
\hat{Y}\ \independent A
\end{equation}

When applied to a binary classifier, independence is often referred to as \emph{demographic parity} (DP), \emph{statistical parity}, \emph{group fairness} or \emph{disparate impact}. This can also be expressed as follows:

\begin{equation}
P(\hat{Y}=1\ |\ A\ =\ a)\ =\ P(\hat{Y}=1\ |\ A\ =\ b),\ \ \ \forall a,b\in\mathcal{A},
\end{equation}

\begin{center}
i.e., for example, the acceptance rate of granting loans to men and women should be equal. 
\end{center}

The ratio of favorable outcomes is also known as positive prediction ratio (PPR) thus independence requires the same PPR across all groups. To have a single number summarizing the magnitude of disparity, it is common to use either the maximum possible difference or minimum possible ratio of PPRs, such that a difference close to 0 or a ratio close to 1 indicates a decision system is “fair” with respect to A and with regards DP. Typically, some tolerance is defined using a threshold which below (or above) a decision is considered “fair enough” and thus acceptable. A commonly cited rule is the “4/5” or “80\% rule” where the selection rate of any SA-defined group should be no less than 80\% of the group with the highest selection rate (i.e., privileged group). This refers to the guidelines of the US Equal Employment Opportunity Commission \citep{Feldman2015}, one of the few examples of a legal framework based on a specific definition of fairness. The crucial factor which differentiates the independence criteria from other metrics is that independence relies only on the distribution of features and decisions, namely $(A,\ X,\ \hat{Y})$, while separation and sufficiency criteria deal with error rate parities, thus make use of the target variable Y as well. Furthermore, independence is referred to as a \emph{non-conservative} measure of fairness \citep{raz2021}, referring to a forced change in the status quo where it is generally not satisfied by the perfect classifier $\hat{Y}=Y$. On the other hand, error rate parities are \emph{conservative} since they trivially hold for the perfect predictor.

\textbf{Separation} strives to ensure equality of errors, or independence of the decision $\hat{Y}$ and sensitive attribute \emph{A} separately for individuals that actually repay their debt and for individuals that do not \citep{Hardt2016}. This is formulated as:

\begin{equation}
\hat{Y}\ \independent A | Y
\end{equation}

In other terms:

\begin{equation}
P(\hat{Y}=1\ |\ A\ =\ a,\ Y\ =\ y)\ =\ P(\hat{Y}=1\ |\ A\ =\ b,\ Y\ =\ y),\ \ \ \forall a,b\in\mathcal{A},\ y\in{0,1}
\end{equation}

\begin{center}
i.e., disparities in groups with different values of A (i.e., male versus female, married versus single) should be completely justified by the value of Y (i.e., repaid or not). 
\end{center}

Separation metrics include \emph{Equality of Odds}, which requires having the same type I (i.e., false positive rate, FPR) and type II (i.e., false negative rate, FNR) error rates across the relevant groups, a more relaxed version known as \emph{Equality of Opportunity}, which requires only equality of FNRs across groups as well as \emph{Predictive Equality}, which requires equality of FPRs across groups. The difference between the latter two metrics is the perspective from which equality is required; \emph{Equality of Opportunity} takes the perspective of people that will repay (i.e., those indeed deserving) whereas \emph{Predictive Equality} takes the perspective of those that will not repay. Depending on the problem at hand, one perspective may be deemed more important than the latter. \citet{Hellman2020} argues that the ratio of false positives and false negatives is a normatively meaningful statistic which should ideally be equalized across the various groups. For example, in a credit lending context, access to credit can be viewed as a positive outcome, however, if an applicant receives a loan which they cannot pay back, it is ultimately harmful for them. This indicates that balancing the risk of FP and FN may be the most pragmatic way forward. Similarly, \citet{Kozodoi2022} argues that separation is a good measure of fairness in credit because it “accounts for the imbalanced misclassification costs of the customer, and, as these imbalanced costs also exist for the financial institution, separation is also able to consider the interests of the loan market." For this reason, we consider \emph{Equality of Odds} in our analysis and separation as our designated fairness criterion in the intersectionality analysis. In sum, separation represents a concept of parity given the ground truth outcome, which considers the point of view of individuals who are subject to the model’s decisions (i.e., considering the number of individuals whose loan request is denied among those who would have repaid), rather than that of the decision-maker. The third concept, sufficiency, considers the opposing view, or parity given the model’s decision (i.e., considering the number of individuals who won’t repay among those given the loan). 

\textbf{Sufficiency} constrains the probability of outcomes to be independent of individual’s SA given their identical non-sensitive information \citep{Kusner2017}. It takes the perspective of people that are given the same decision (i.e., approved for a loan) and requires parity among them irrespective of their SA. A fairness metric which focuses on this type of error rate is called \emph{Predictive Parity} and noted as follows:

\begin{equation}
P(Y=1\ |\ A\ =\ a,\ \hat{Y}\ =\ 1)\ =\ P(Y=1\ |\ A\ =\ b,\ \hat{Y}\ =\ 1),\ \ \ \forall a,b\in\mathcal{A}
\end{equation}

\begin{center}
i.e., the model should have equal precision across the sensitive groups. When requiring this condition to also hold for the case Y= 0, the conditional independence statement is noted as:
\end{center}

\begin{equation}
Y\ \independent A | \hat{Y}
\end{equation}

In general, fairness metrics can be contradictory or complimentary depending on the application and the data \citep{corbettdavies2018}. We note that in cases such as credit lending where the target variable Y represents the decision of the algorithm or loan officer, separation and sufficiency-based criteria must be used with care because Y itself can be prone to bias for certain groups. Even if Y represents repayment of a loan, a form of selection bias is likely already at work. For example, the model only has information on applicants who received a loan in the first place (i.e., historical cases), and these are likely not representative of the whole population of applicants which includes those originally denied \citep{Crook2004}.  

Our fairness analysis considers all three criteria using a differential nature. This is measured based on the difference between the unprivileged group and privileged group, thus zero represents the value at which a model would be considered unbiased. For example, \emph{Statistical Parity Difference} (SPD) measures the difference in the probability of being labelled with the favorable outcome (i.e., approved) between an individual that belongs to the unprivileged group and an individual that belongs to the privileged group. It is calculated with the following equation: 

\begin{equation}
SPD=\ P(\hat{Y}=1\ |\ A\ =\ unprivileged)-\ P(\hat{Y}=1\ |\ A\ =\ privileged)
\end{equation}

This is similarly done for the separation criterion using Average Odds Difference (AOD) where a value of 0 indicates equality of odds \citep{ibmmetrics}:

\begin{equation}
AOD={\frac{1}{2}(FPR}_{A=unprivileged}-{FPR}_{A=privileged})+\ {(TPR}_{A=unprivileged}-{TPR}_{A=privileged})
\end{equation}

And the sufficiency criterion uses \emph{Predictive Rate Parity} (PRP) by ensuring minimal differences in the positive predictive value (PPV, i.e., precision) across subgroups:

\begin{equation}
PRP={PPV}_{A=unprivileged}-{PPV}_{A=privileged}
\end{equation}

The level of fairness is assessed based on the threshold-based conditions defined in Table \ref{table:table3}, following the methodology of \citet{Stevens2020}. 

\begin{table}[h]
\includegraphics[width=8cm]{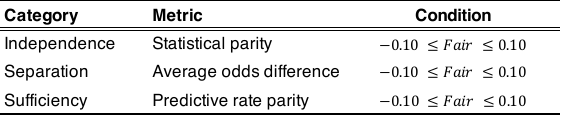}
\centering
\caption{Three fairness metrics and their conditions. }
\label{table:table3}
\end{table}

\section{Empirical Analysis and Results}
\label{sec:Results}
\subsection{Interpretability of Baseline Model}

\begin{figure}[h]
\includegraphics[width=14cm]{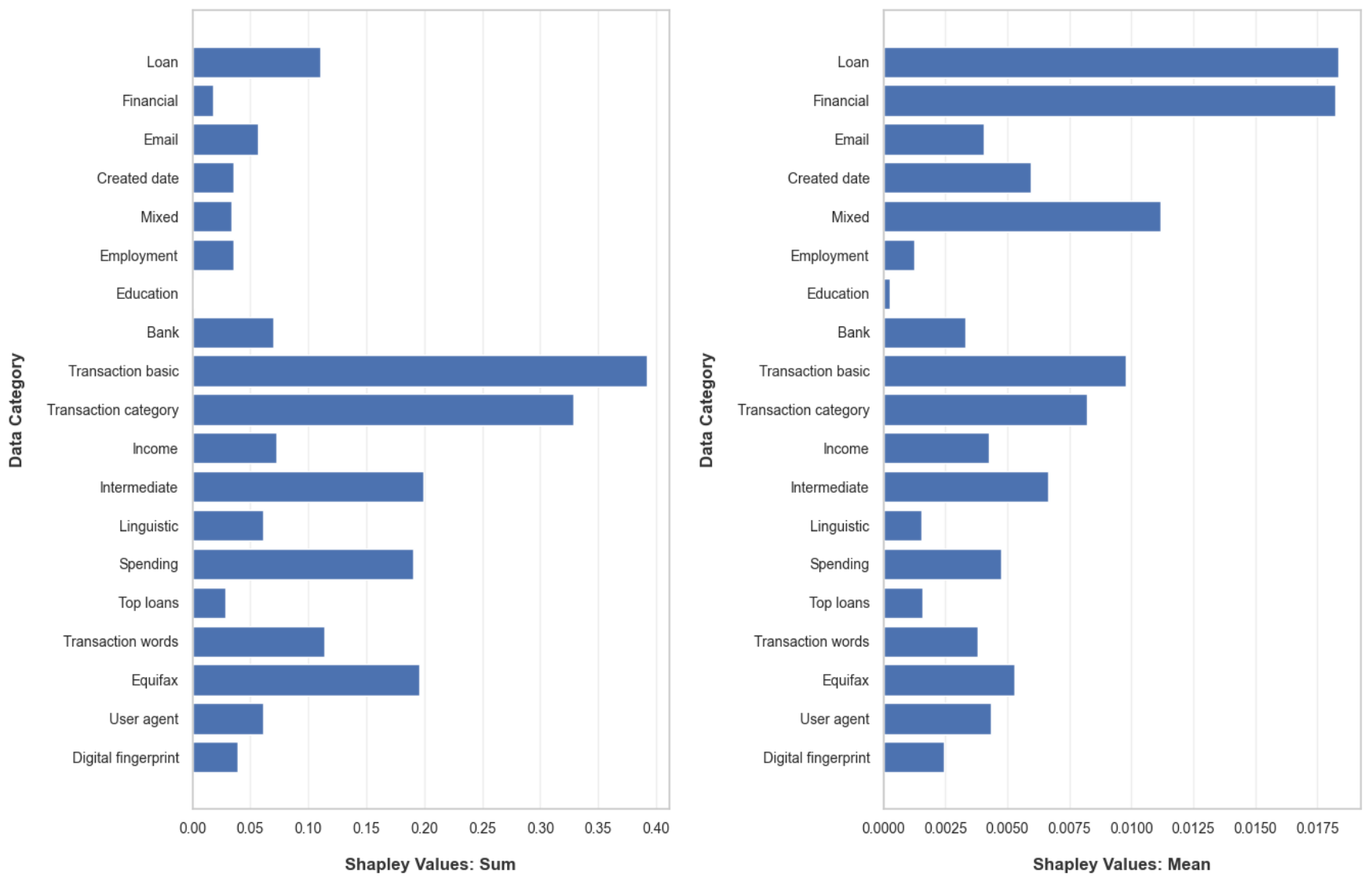}
\centering
\caption{Right displays Shapley values by sum, left displays Shapley values by mean, per feature category.}
\label{fig:fig2}
\end{figure}

To understand the importance of data categories as a whole while accounting for the high number of features, we first aggregate the Shapley values across data categories, instead of looking at each feature individually. To account for the different data types, we sum and average the resulting Shapley values for the baseline model by category as listed in Table \ref{table:table2}. The former informs us of which group contains the more important features, however, naturally the larger groups will have greater importance. To compensate for this, the mean is also used to derive which group as a whole is more important. Figure \ref{fig:fig2} (left) displays the data categories by sum. We find that the data categories ‘transaction basic’ and ‘transaction category’ play the most prominent role, which represent transaction-based features (e.g., average transactions per day, min/max balance, number of grocery or hospital-related transactions, etc.). This is likely because bank transaction records, i.e., what you buy, how often you buy, how consistently you buy, are reflections of consumer risk characteristics based on financial habits and preferences \citep{Djeundje2021, Thompson2020, Tobback2019}. Figure \ref{fig:fig2} (right) displays the data categories by averages. Categories ‘loan’ and ‘financial’ are overall weighted higher in average, referring to features related to loan amount, loan amount to income, and financial features such as one’s rent and stated income. This is reasonable as these features typically make up an applicant’s current financial profile and are commonly used in traditional credit scoring models to determine creditworthiness \citep{Mester1997}. This interpretability analysis lays the groundwork for determining feature importance by category, which is crucial in our next analysis when testing different data-based variations of our model.

\subsection{'High-level' Fairness}

For the baseline model, where all original data types remain as inputs, the fairness results are depicted in Table \ref{table:table4}. Results which satisfy the fairness conditions as stated in Table \ref{table:table3} are boldfaced. The baseline model obtained an accuracy of 0.7041, a F1 score of 0.7902 and an AUC score of 0.7327. We find that the fairness conditions are satisfied with minimal disparity for the majority of SAs when observed individually. The highest disparity is observed when comparing those with no children and multiple children (independence disparity is -0.03, separation disparity is -0.031, sufficiency disparity is 0.008) as well as married compared to divorced individuals (independence disparity is -0.082, separation disparity is -0.066, sufficiency disparity is -0.023), however, they do not surpass our conditions for unfairness. Interestingly, while previous empirical works have often highlighted gender disparity in fairness outcomes for credit datasets \citep{Kim2020}, we see that for our dataset, gender holds the least magnitude of disparity. 

\begin{table}[h]
\includegraphics[width=\linewidth]{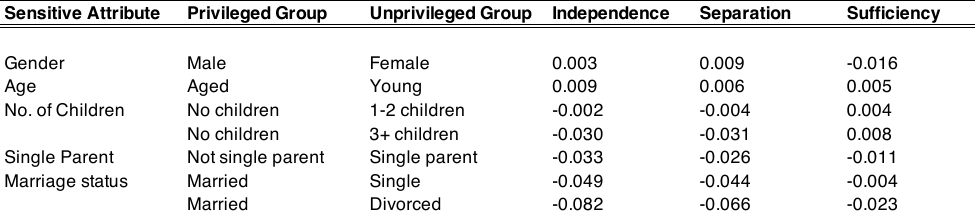}
\centering
\caption{Fairness results for baseline model, across all sensitive attributes and three fairness criteria.}
\label{table:table4}
\end{table}

The authors feel this may be attributed to the domain of our study, microfinance lending, which is a type of credit product that commends itself with having a successful financial \emph{and} philanthropic model \citep{Lee2014}. Microfinance institutions (MFI) are often advertised as a suitable alternative for individuals who are typically unable to access loans from traditional financial institutions, due to their financial capability or lack of sufficient credit history. MFIs reflect the practice of predominately extending loans to poor or low-income individuals and have proven capable of reaching individuals that previously lacked access to financial services altogether, such as women in rural areas \citep{Hermes2018}. Therefore, we have in this study used a context that is challenging and scarce in the current fairness literature. MF is unique in that it often targets the underbanked in the first place, thereby diverging from more typical patterns of disparity seen in traditional loan products (e.g., mortgages or consumer loans) which target the general population. In a way, this results in a tendency towards reversed disparity or preference-based bias \citep{Hu2000}; for example, funding more women as part of their philanthropic aim, which is supported by our HL fairness results. However, when observed in greater depth, we find that even in this context we observe noticeable discrimination but in a different setting, which is detailed next. 

We extend this analysis to include the data-based variations of the model: (1) traditional, (2) traditional \& alternative credit (AC), and traditional \& digital footprint (DF) with the goal of determining whether similar trends can be observed when certain data types are excluded in the analysis. The performance metrics of each model can be found in the Appendix (Table \ref{table:apx3}). The fairness results are displayed in Table \ref{table:table5} and those that satisfy fairness conditions are marked in boldfaced. An asterisk indicates significant p-value (<0.05) reported for the t-test comparing the Baseline model to each of the three model variations. The results have several implications. First, we see that as the amount of information through feature groups increases, the predictive ability of the test set also increases which supports previous works in alternative credit literature \citep{dimaggio2022, Lu2019, oskardottir2019}. The baseline model, which includes all features, dominates with regards to predictability and is therefore used as the most suitable use case in the next intersectionality analysis. Second, we find that the traditional model results in the greatest magnitude of change in higher level fairness outcomes, with the traditional \& AC and traditional \& DF variations falling in the middle. In particular, there is an exacerbation of disparity for already existing cases; for example, the \emph{independence} value comparing married and divorced individuals is found to be -0.082. However, with the traditional model, this disparity is decreased further by -0.039, resulting in a -0.121* disparity in the traditional model which falls outside of our fairness conditions (i.e., significantly unfair). This similarly occurs for the \emph{separation} criterion, indicating that traditional credit scoring approaches may result in greater disparity in prediction results for SA-defined groups. However, the majority of fairness outcomes, regardless of model type, majorly fulfill our conditions indicating that higher-level fairness still holds in most of these cases. 

\begin{table}[h]
\includegraphics[width=\linewidth]{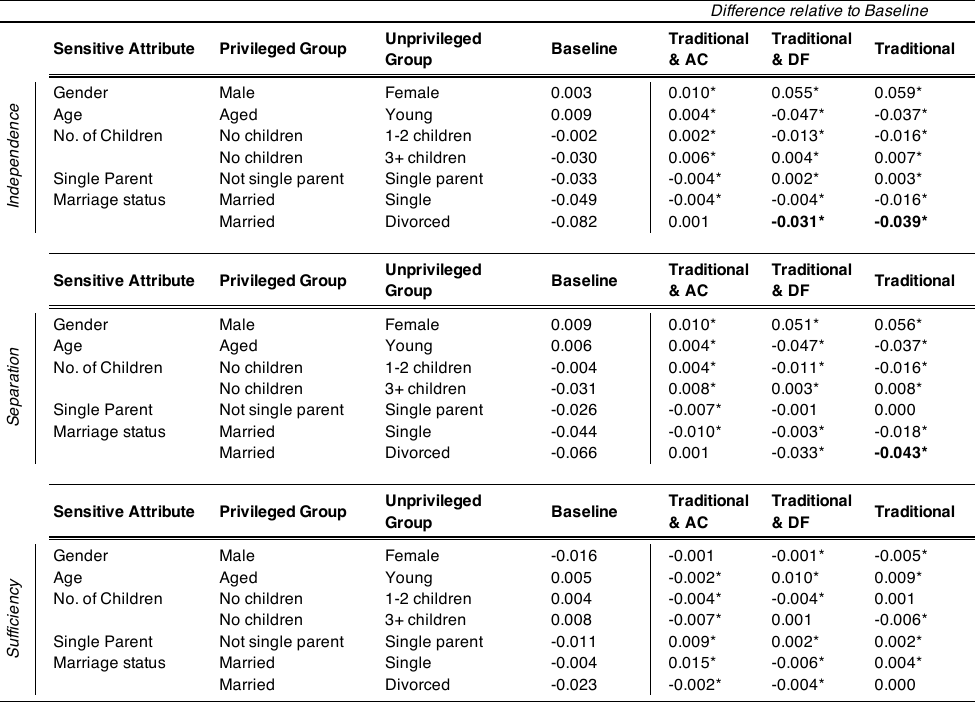}
\centering
\caption{Fairness results for data-based variations of the baseline model. \emph{Note: Values represent magnitude differences relative to baseline model; positive values indicate improvement in fairness and negative values indicate worsening in fairness.}}
\label{table:table5}
\end{table}

\subsection{'Low-level' Fairness}

We extend our investigation to comprise intersectionality next. The results can be found in Table \ref{table:table6}, where deeper subgroups consisting of two-SA combinations are considered (e.g., young female, single parent with 3+ children). We term this “lower-level” fairness, referencing the deeper subgroups created by multiple and multinomial SAs simultaneously. As discussed in Section \ref{sec:Related Work}, individuals rarely fall under one SA-defined group at a time, but rather individuals often experience multiple identities and consequently the associated societal stereotypes, expectations and restrictions simultaneously. In this analysis, we observe an increasing number of cases which do not comply with the fairness conditions laid out in Table \ref{table:table3}, which are highlighted in boldface. 

\begin{table}
\includegraphics[width=11cm]{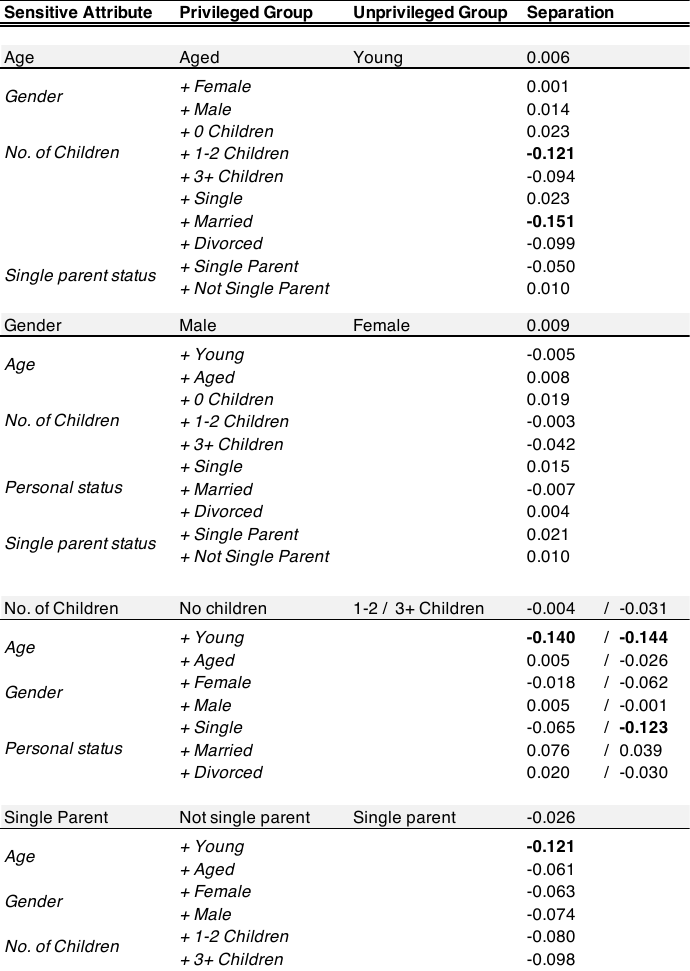}
\centering
\caption{Fairness results for intersectionality combinations using the baseline model. \emph{Note: Values represent magnitude differences relative to baseline model; positive values indicate improvement in fairness.}}
\label{table:table6}
\end{table}

For example, when looking at the attribute number of children (with no children being the privileged group), the SA individually remains compliant to fairness conditions with a separation disparity of -0.004 and -0.031, respectively for 1-2 children and 3+ children. However, when this attribute interacts with age (i.e., being young), we see the disparity exacerbate to -0.140 and -0.144. This is similarly the case with age in relation to marriage status, with married young individuals suffering greater disparity at -0.151 compared to aged, married individuals. Additionally, when comparing young individuals, young single parents are also at a disadvantage when compared to young individuals who are not single parents (-0.121). Overall, we see that being a young individual, in combination with being a single parent and having children fare worse in predictive models for repayment compared to other groups. This may also be due to factors such as attributed lack of financial stability, greater financial responsibilities, not having yet finished their education, etc. Furthermore, the greater the number of children, the worse the model performs (i.e., higher error rates); this is a trend seen in combination with any of the target SAs (e.g., gender, age, or marital status). However, this was not as blatant when investigating outcomes in relation to higher-level fairness previously.

\begin{figure}[h]
\includegraphics[width=14cm]{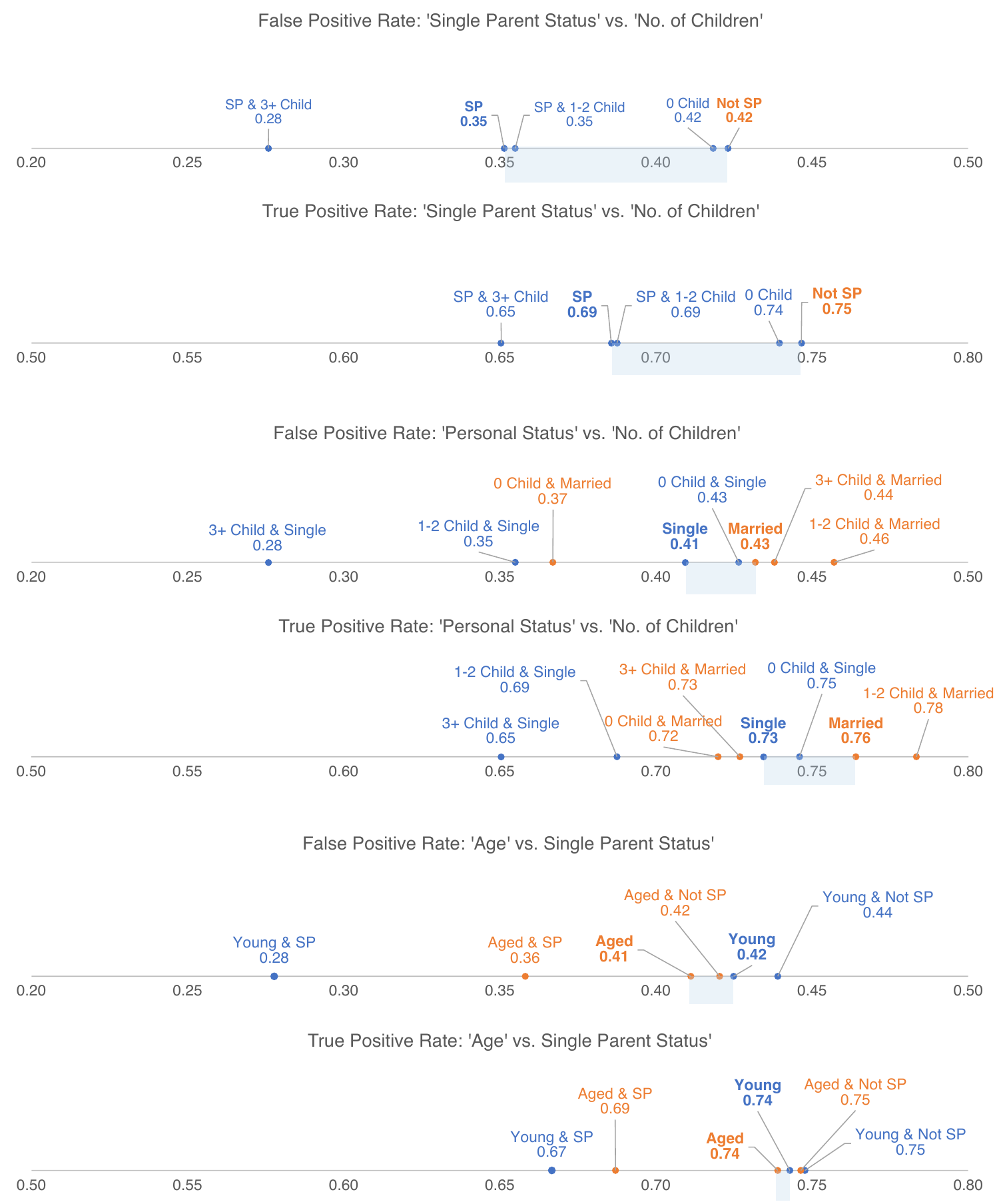}
\centering
\caption{Visualization of intersectionality-based fairness, separated by components of Separation criterion.}
\label{fig:fig3}
\end{figure}

The amplification of subgroup-based errors becomes even more evident in the visualization presented in Figure \ref{fig:fig3}, which breaks down the separation criterion into its components: FPR and TPR. This is displayed for the intersectional comparisons between: (1) ‘single parent’ and ‘no. of children’, (2) ‘personal status’ and ‘no. of children’ and (3) ‘age’ and ‘single parent status.’ The higher-level fairness results (i.e., single or not single status) are indicated in boldface with disparity highlighted in blue and the intersectional combinations are indicated in orange and blue based on privileged or unprivileged group status, respectively. In the case of FPR versus TPR, the FPR represents the proportion of applicants who \emph{actually} default that were incorrectly granted loans. TPR, on the other hand, represents the proportion of applicants who \emph{actually} pay back loans that were accurately granted loans. It is problematic if either of these rates are significantly higher for the privileged group, which we observe in intersectionality cases. The narrow disparity (i.e., distance) between groups in boldface indicate the higher-level fairness outcomes, which masks the deeper problem. Rather, it averages and thus underestimates the wider disparity between its components; the subgroups displayed in orange versus blue show greater disparity indicating an exacerbation of the fairness problem at deeper levels when high-level groupings are broken down. 

\section{Discussion}
\label{sec:Discussion}

The case study presented in this paper shows that our test design is relevant in the context of microcredit when assessing the fairness of the loan granting process with respect to pre-defined sensitive groups. Overall, our study suggests that the mode of access to microcredit in the Spanish market may work in non-equalizing manner, largely maintaining fair outcomes overall but falling short with inequities of accessing credit for those who identify with multiple, underprivileged groups. The study employed an intersectional approach to assess horizontal inequity by sensitive attributes including gender, age, single parent status, number of children, and personal status in microcredit lending. The results suggest that whereas on a higher-level, fairness outcomes may seemingly comply with fair conditions when viewed along and across single axis conditions; however, when two SAs are combined simultaneously to reflect the more realistic multiplicity of identities (which individuals hold in daily life), the outcomes show amplifying disparity and fail to meet fair conditions. By looking at error rates in terms of fraction of errors over the ground truth (i.e., separation criterion), the study found that the impact of being a young individual with children, young single individual and young single parent compounded with each other and thus put them at a greater disadvantage compared to their privileged counterparts. In other words, these groups suffer proportionally higher number of individuals whose loan request is denied among those who would have repaid (had they been given a loan), thus limiting their credit opportunities and ultimately, resulting in an unjust distribution of crucial liberal goods. We conclude that the core problem may be more complicated than current literature demonstrates, as it mainly focuses on single axis thinking. 

We note the wider implications of this work. In particular, young and single individuals as well as those with multiple children are typically financially challenged populations \citep{Kim2023}. They tend to experience some combination of decreased income, greater financial burdens, schedule changes, and higher job loss risks and are more likely to engage in jobs with less stability, income, hours, and access to benefits such as childcare, health insurance or paid leave \citep{Morris2004, Scott2005}. This is compounded with overall less familial support, which can typically help alleviate these liabilities. Thus, the individuals most affected also have the least resources to shoulder financial burdens and may have a greater need for small sums of credit to temporarily overcome these financial burdens, without having to resort to payday lenders and the exorbitant interest rates associated with them. Furthermore, classist and sexist policies and norms tend to exacerbate inequities and constrain employment opportunities and financial pathways that compound over time \citep{Bailey2017, Bell2000, Zambrana2017}. This can exacerbate later-life inequities relating to financial status and stability, making their well-being increasingly precarious. We call attention to these risks groups by highlighting the nature of intersecting SAs and how they can reinforce each other to further amplify the harms of complex structural inequities in our society. 

Those who identify with multiple SAs are, however, a niche part of the overall population, as shown in the Appendix. One may therefore question the external validity of our results. We contend that, although individuals who fall under multiple groupings are significantly smaller in population size compared to higher-level groupings, it remains a substantial number; the majority of subgroups remain larger than 1000 instances.  We reference the popularly used German Credit dataset \citep{UCI} which holds only 1000 instances yet has been commonly sourced as a benchmarking dataset in the fair ML literature to draw significant conclusions. Furthermore, our intersectional discrimination approach is transposable to any type of credit product and therefore applicable to a wide range of loan providers including for-profit and prosocial FinTech lenders. The study also illustrates how enhancement of traditional credit models with alternative credit data can also result in complex feature relationships and modeling processes which result in varied fairness outcomes not captured when using traditional credit assessment methods. FinTech alternative credit lenders and researchers therefore need to pay greater attention to intersectional inequities that can be perceived on a higher level to be minimal, and targeting subgroups of individuals who may be experience an amplification of prediction error-based harm when accessing microcredit. 

While discriminatory behaviors can have significant negative economic consequences, their most serious failing is ethical. Discrimination in lending on the basis of sociodemographic information (e.g., age, gender, ethnic origin or religion) is prohibited by law in many countries, including the Equal Credit Opportunity Act in the US and Article 13 of the European Commission Treaty in the European Union. In accordance with these standards, none of the SAs which are considered in our analysis were used to train the classifiers. We note two major drawbacks to these regulatory constraints. Firstly, they have failed to keep up with the technological advancements and increasingly intrusive data being gathered by companies \citep{Wachter2021}. For example, while legally protected attributes are not directly used as inputs into the model, seemingly trivial features such as zip code, income, education level, school attendance, can serve as proxies for structural racism, sexism, classism, etc. \citep{Custers2013}. Whether intentional or unintentional, they risk similar discriminatory outcomes. Secondly, law protects only those who fall under a limited range of protected characteristics; however, this lends to a major loophole. It fails to clarify how the rule is to be implemented when it comes to intersectional biases. For example, is discrimination against Black women unlawful if both Black men and White women are treated fairly? Second, legislation fails to consider the additional harm posed by attributes we deem “sensitive,” as an extension to legally protected characteristics which we have highlighted in this study. In a technological environment of increasingly granular, personalized insights, there is a major risk of identifying more at-risk groups of individuals which was less of a problem in previous traditional data settings. We have, in our study, brought these concerns to light with an empirical analysis of a real-world microfinance case. The dataset utilized is a product of the increasingly granular data trend we are seeing across credit services, also known as alternative credit data. We have also considered non-standard SAs and their combination and found that profiling via the proxying of SAs is also observed in the microfinance context as evidenced by the (un)fairness results albeit materializing in different form from traditional credit settings. 

\subsection{Conclusion}

Our examination of demographic profiling and predictive analytics in automated credit decision-making looks to engage more directly with specific intersections of data use and historically situated discrimination. Given the increasing scale and scope of credit market operations alongside technological advancements, greater clarification on this issue would be welcome. Discrimination and ensuring fairness in lending continues to be an issue, even with regulation in place albeit outdated. Therefore, we call for relevant tools to further investigate the presence of unfairness in ADM systems. Moreover, greater attention needs to be paid to fairness in ML models which adequately captures the needs of those who identify with multiple group identities when estimating horizontal equity in credit decision-making. Our work presents a step in this direction.

\paragraph{Acknowledgements}
The authors are grateful to the data provider who has chosen to remain anonymous. S. Kim is grateful to Baillie Gifford for funding. 

\graphicspath{ {./images/} }

\newpage
\bibliographystyle{unsrtnat}
\bibliography{references}  

\begin{thebibliography}{115}
\providecommand{\natexlab}[1]{#1}
\providecommand{\url}[1]{\texttt{#1}}
\expandafter\ifx\csname urlstyle\endcsname\relax
  \providecommand{\doi}[1]{doi: #1}\else
  \providecommand{\doi}{doi: \begingroup \urlstyle{rm}\Url}\fi

\bibitem[Smith et~al.(2016)Smith, Patil, and Muñoz]{Smith2016}
Megan Smith, DJ~Patil, and Cecilia Muñoz.
\newblock Big risks, big opportunities: the intersection of big data and civil
  rights | whitehouse.gov, 5 2016.
\newblock URL
  \url{https://obamawhitehouse.archives.gov/blog/2016/05/04/big-risks-big-opportunities-intersection-big-data-and-civil-rights}.

\bibitem[Boyd and Crawford(2012)]{Boyd2012}
Danah Boyd and Kate Crawford.
\newblock Critical questions for big data.
\newblock \emph{Information, Communication \& Society}, 15:\penalty0 662--679,
  6 2012.
\newblock ISSN 1369118X.
\newblock \doi{10.1080/1369118X.2012.678878}.
\newblock URL
  \url{https://www.tandfonline.com/doi/abs/10.1080/1369118X.2012.678878}.

\bibitem[Einav et~al.(2013)Einav, Jenkins, and Levin]{Einav2013}
Liran Einav, Mark Jenkins, and Jonathan Levin.
\newblock The impact of credit scoring on consumer lending.
\newblock \emph{Journal of Economics}, 44:\penalty0 249--274, 2013.
\newblock \doi{10.1111/1756-2171.12019}.

\bibitem[Agarwal et~al.(2019)Agarwal, Alok, Ghosh, and Gupta]{Agarwal2019}
Sumit Agarwal, Shashwat Alok, Pulak Ghosh, and Sudip Gupta.
\newblock Financial inclusion and alternate credit scoring: Role of big data
  and machine learning in fintech.
\newblock \emph{Indian School of Business}, 12 2019.

\bibitem[Brevoort et~al.(2016)Brevoort, Grimm, and Kambara]{Brevoort2016}
Kenneth~P Brevoort, Philipp Grimm, and Michelle Kambara.
\newblock Credit invisibles and the unscored.
\newblock \emph{Cityscape}, 18, 2016.

\bibitem[Bhutta et~al.(2015)Bhutta, Skiba, and Tobacman]{Bhutta2015}
Neil Bhutta, Paige~Marta Skiba, and Jeremy Tobacman.
\newblock Payday loan choices and consequences.
\newblock \emph{Journal of Money, Credit and Banking}, 47:\penalty0 223--260, 3
  2015.
\newblock ISSN 1538-4616.
\newblock \doi{10.1111/JMCB.12175}.
\newblock URL \url{https://onlinelibrary.wiley.com/doi/full/10.1111/jmcb.12175
  https://onlinelibrary.wiley.com/doi/abs/10.1111/jmcb.12175
  https://onlinelibrary.wiley.com/doi/10.1111/jmcb.12175}.

\bibitem[Hacker(2018)]{Hacker2018}
Philipp Hacker.
\newblock Teaching fairness to artificial intelligence: Existing and novel
  strategies against algorithmic discrimination under eu law.
\newblock \emph{Common Market Law Review}, 55, 8 2018.

\bibitem[Langenbucher(2020)]{Langenbucher2020}
Katja Langenbucher.
\newblock Responsible a.i.-based credit scoring – a legal framework.
\newblock \emph{European Business Law Review}, 31, 8 2020.

\bibitem[Purificato et~al.(2022)Purificato, Lorenzo, Fallucchi, and
  Luca]{Purificato2022}
Erasmo Purificato, Flavio Lorenzo, Francesca Fallucchi, and Ernesto William~De
  Luca.
\newblock The use of responsible artificial intelligence techniques in the
  context of loan approval processes.
\newblock \emph{International Journal of Human–Computer Interaction}, 2022.
\newblock ISSN 15327590.
\newblock \doi{10.1080/10447318.2022.2081284}.
\newblock URL
  \url{https://www.tandfonline.com/doi/abs/10.1080/10447318.2022.2081284}.

\bibitem[Wachter et~al.(2021)Wachter, Mittelstadt, and Russell]{Wachter2021}
Sandra Wachter, Brent Mittelstadt, and Chris Russell.
\newblock Why fairness cannot be automated: Bridging the gap between eu
  non-discrimination law and ai.
\newblock \emph{Computer Law \& Security Review}, 41:\penalty0 105567, 7 2021.
\newblock ISSN 0267-3649.
\newblock \doi{10.1016/J.CLSR.2021.105567}.

\bibitem[Heikkiläarchive(2022)]{Heikkilaarchive2022}
Melissa Heikkiläarchive.
\newblock The eu wants to put companies on the hook for harmful ai.
\newblock \emph{MIT Technology Review}, 2022.
\newblock URL
  \url{https://www.technologyreview.com/2022/10/01/1060539/eu-tech-policy-harmful-ai-liability/}.

\bibitem[Aitken(2017)]{Aitken2017}
Rob Aitken.
\newblock ‘all data is credit data’: Constituting the unbanked.
\newblock \emph{Competition and Change}, 21:\penalty0 274--300, 6 2017.
\newblock ISSN 14772221.
\newblock \doi{10.1177/1024529417712830}.
\newblock URL
  \url{https://journals.sagepub.com/doi/abs/10.1177/1024529417712830}.

\bibitem[Tobback and Martens(2019)]{Tobback2019}
Ellen Tobback and David Martens.
\newblock Retail credit scoring using fine-grained payment data.
\newblock \emph{Journal of the Royal Statistical Society: Series A (Statistics
  in Society)}, 182:\penalty0 1227--1246, 10 2019.

\bibitem[Shmueli(2016)]{Shmueli2016}
Galit Shmueli.
\newblock Analyzing behavioral big data: Methodological, practical, ethical,
  and moral issues.
\newblock \emph{Quality Engineering}, 29:\penalty0 57--74, 1 2016.
\newblock ISSN 15324222.

\bibitem[Thompson et~al.(2020)Thompson, Feng, Reesor, and Grace]{Thompson2020}
John R.~J. Thompson, Longlong Feng, R.~Mark Reesor, and Chuck Grace.
\newblock Know your clients' behaviours: A cluster analysis of financial
  transactions.
\newblock \emph{Journal of Risk and Financial Management}, 14:\penalty0 50, 5
  2020.

\bibitem[Vissing-Jorgensen(2012)]{vissing2012}
Annette Vissing-Jorgensen.
\newblock Consumer credit: Learning your customer's default risk from what
  (s)he buys.
\newblock \emph{SSRN Electronic Journal}, 2012.

\bibitem[Omarini(2018)]{Omarini2018}
Anna~Eugenia Omarini.
\newblock Banks and fintechs: How to develop a digital open banking approach
  for the bank's future.
\newblock \emph{International Business Research}, 11:\penalty0 23--36, 2018.

\bibitem[Remolina(2019)]{Remolina2019}
Nydia Remolina.
\newblock Open banking: Regulatory challenges for a new form of financial
  intermediation in a data-driven world.
\newblock \emph{SMU Centre for AI \& Data Governance Research}, 5, 2019.

\bibitem[Foos et~al.(2010)Foos, Norden, Weber, Foos, Norden, and
  Weber]{Foos2010}
Daniel Foos, Lars Norden, Martin Weber, Daniel Foos, Lars Norden, and Martin
  Weber.
\newblock Loan growth and riskiness of banks.
\newblock \emph{Journal of Banking \& Finance}, 34:\penalty0 2929--2940, 2010.
\newblock URL
  \url{https://econpapers.repec.org/RePEc:eee:jbfina:v:34:y:2010:i:12:p:2929-2940}.

\bibitem[Khandani et~al.(2010)Khandani, Kim, and Lo]{Khandani2010}
Amir~E. Khandani, Adlar~J. Kim, and Andrew~W. Lo.
\newblock Consumer credit-risk models via machine-learning algorithms.
\newblock \emph{Journal of Banking \& Finance}, 34:\penalty0 2767--2787, 11
  2010.
\newblock ISSN 0378-4266.
\newblock \doi{10.1016/J.JBANKFIN.2010.06.001}.

\bibitem[Bellotti and Crook(2013)]{Bellotti2013}
Tony Bellotti and Jonathan Crook.
\newblock Forecasting and stress testing credit card default using dynamic
  models.
\newblock \emph{International Journal of Forecasting}, 29:\penalty0 563--574,
  10 2013.
\newblock ISSN 0169-2070.
\newblock \doi{10.1016/J.IJFORECAST.2013.04.003}.

\bibitem[Lee et~al.(2022)Lee, Yang, and Anderson]{Lee2022}
Jung~Youn Lee, Joonhyuk Yang, and Eric Anderson.
\newblock Banking the unbanked: Using grocery data for credit decisions.
\newblock \emph{Wharton Customer Analytics Research Paper}, 8 2022.
\newblock \doi{10.2139/SSRN.3868547}.
\newblock URL \url{https://papers.ssrn.com/abstract=3868547}.

\bibitem[Berg et~al.(2020)Berg, Burg, Gombović, Puri, and Karolyi]{Berg2020}
Tobias Berg, Valentin Burg, Ana Gombović, Manju Puri, and Andrew Karolyi.
\newblock On the rise of fintechs: Credit scoring using digital footprints.
\newblock \emph{Review of Financial Studies}, 33:\penalty0 2845--2897, 11 2020.

\bibitem[Pedro et~al.(2015)Pedro, Proserpio, and Oliver]{Pedro2015}
Jose~San Pedro, Davide Proserpio, and Nuria Oliver.
\newblock Mobiscore: Towards universal credit scoring from mobile phone data.
\newblock \emph{Lecture Notes in Computer Science (including subseries Lecture
  Notes in Artificial Intelligence and Lecture Notes in Bioinformatics)},
  9146:\penalty0 195--207, 2015.
\newblock ISSN 16113349.
\newblock \doi{10.1007/978-3-319-20267-9_16/COVER}.
\newblock URL
  \url{https://link.springer.com/chapter/10.1007/978-3-319-20267-9_16}.

\bibitem[Björkegren and Grissen(2020)]{bjorkegren2020}
Daniel Björkegren and Darrell Grissen.
\newblock Behavior revealed in mobile phone usage predicts credit repayment.
\newblock \emph{The World Bank Economic Review}, 34:\penalty0 618--634, 10
  2020.
\newblock ISSN 0258-6770.
\newblock \doi{10.1093/WBER/LHZ006}.

\bibitem[Lin et~al.(2012)Lin, Prabhala, and Viswanathan]{Lin2012}
Mingfeng Lin, Nagpurnanand~R. Prabhala, and Siva Viswanathan.
\newblock Judging borrowers by the company they keep: Friendship networks and
  information asymmetry in online peer-to-peer lending.
\newblock \emph{https://doi.org/10.1287/mnsc.1120.1560}, 59:\penalty0 17--35, 9
  2012.
\newblock ISSN 00251909.
\newblock \doi{10.1287/MNSC.1120.1560}.
\newblock URL
  \url{https://pubsonline.informs.org/doi/abs/10.1287/mnsc.1120.1560}.

\bibitem[Zhang et~al.(2016)Zhang, Jia, Diao, Hai, and Li]{Zhang2016}
Yuejin Zhang, Hengyue Jia, Yunfei Diao, Mo~Hai, and Haifeng Li.
\newblock Research on credit scoring by fusing social media information in
  online peer-to-peer lending.
\newblock \emph{Procedia Computer Science}, 91:\penalty0 168--174, 2016.

\bibitem[Baesens et~al.(2003)Baesens, Gestel, Viaene, Stepanova, Suykens, and
  Vanthienen]{Baesens2003}
B.~Baesens, T.~Van Gestel, S.~Viaene, M.~Stepanova, J.~Suykens, and
  J.~Vanthienen.
\newblock Benchmarking state-of-the-art classification algorithms for credit
  scoring.
\newblock \emph{Journal of the Operational Research Society}, 54:\penalty0
  627--635, 2003.

\bibitem[Fuster et~al.(2019)Fuster, Plosser, Schnabl, and Vickery]{Fuster2019}
Andreas Fuster, Matthew Plosser, Philipp Schnabl, and James Vickery.
\newblock The role of technology in mortgage lending.
\newblock \emph{The Review of Financial Studies}, 32:\penalty0 1854--1899, 5
  2019.
\newblock ISSN 0893-9454.
\newblock \doi{10.1093/RFS/HHZ018}.
\newblock URL \url{https://academic.oup.com/rfs/article/32/5/1854/5427780}.

\bibitem[Lessmann et~al.(2015)Lessmann, Baesens, Seow, and
  Thomas]{Lessmann2015}
Stefan Lessmann, Bart Baesens, Hsin~Vonn Seow, and Lyn~C. Thomas.
\newblock Benchmarking state-of-the-art classification algorithms for credit
  scoring: An update of research.
\newblock \emph{European Journal of Operational Research}, 247:\penalty0
  124--136, 2015.

\bibitem[Morse(2015)]{Morse2015}
Adair Morse.
\newblock Peer-to-peer crowdfunding: Information and the potential for
  disruption in consumer lending.
\newblock \emph{Annual Review of Financial Economics}, 7:\penalty0 463--482, 2
  2015.
\newblock \doi{10.3386/W20899}.
\newblock URL \url{https://www.nber.org/papers/w20899}.

\bibitem[de~Castro~Vieira et~al.(2019)de~Castro~Vieira, Barboza, Sobreiro, and
  Kimura]{vieira2019}
José~Rômulo de~Castro~Vieira, Flavio Barboza, Vinicius~Amorim Sobreiro, and
  Herbert Kimura.
\newblock Machine learning models for credit analysis improvements: Predicting
  low-income families’ default.
\newblock \emph{Applied Soft Computing}, 83:\penalty0 105640, 10 2019.
\newblock ISSN 1568-4946.
\newblock \doi{10.1016/J.ASOC.2019.105640}.

\bibitem[Kruppa et~al.(2013)Kruppa, Schwarz, Arminger, and Ziegler]{Kruppa2013}
Jochen Kruppa, Alexandra Schwarz, Gerhard Arminger, and Andreas Ziegler.
\newblock Consumer credit risk: Individual probability estimates using machine
  learning.
\newblock \emph{Expert Systems with Applications}, 40:\penalty0 5125--5131,
  2013.

\bibitem[Malagon et~al.(2022)Malagon, Troncoso, Rubio, and Ponce]{Malagon2022}
Emmanuel Malagon, Daniel Troncoso, Andres Rubio, and Hiram Ponce.
\newblock Machine learning techniques in credit default prediction.
\newblock \emph{Lecture Notes in Computer Science (including subseries Lecture
  Notes in Artificial Intelligence and Lecture Notes in Bioinformatics)}, 13612
  LNAI:\penalty0 204--211, 2022.
\newblock ISSN 16113349.
\newblock \doi{10.1007/978-3-031-19493-1_17/FIGURES/3}.
\newblock URL
  \url{https://link.springer.com/chapter/10.1007/978-3-031-19493-1_17}.

\bibitem[Agier and Szafarz(2013)]{Agier2013_mfandgender}
Isabelle Agier and Ariane Szafarz.
\newblock Microfinance and gender: Is there a glass ceiling on loan size?
\newblock \emph{World Development}, 42:\penalty0 165--181, 2 2013.
\newblock ISSN 0305-750X.
\newblock \doi{10.1016/J.WORLDDEV.2012.06.016}.

\bibitem[Bayer et~al.(2018)Bayer, Ferreira, and Ross]{Bayer2018}
Patrick Bayer, Fernando Ferreira, and Stephen~L. Ross.
\newblock What drives racial and ethnic differences in high-cost mortgages? the
  role of high-risk lenders.
\newblock \emph{The Review of Financial Studies}, 31:\penalty0 175--205, 1
  2018.
\newblock ISSN 0893-9454.
\newblock \doi{10.1093/RFS/HHX035}.
\newblock URL \url{https://academic.oup.com/rfs/article/31/1/175/3782656}.

\bibitem[Berkovec et~al.(1998)Berkovec, Canner, Gabriel, and
  Hannan]{Berkovec1998}
James~A. Berkovec, Glenn~B. Canner, Stuart~A. Gabriel, and Timothy~H. Hannan.
\newblock Discrimination, competition, and loan performance in fha mortgage
  lending.
\newblock \emph{The Review of Economics and Statistics}, 80:\penalty0 241--250,
  5 1998.
\newblock ISSN 0034-6535.
\newblock \doi{10.1162/003465398557483}.
\newblock URL
  \url{https://direct.mit.edu/rest/article/80/2/241/57067/Discrimination-Competition-and-Loan-Performance-in}.

\bibitem[Black et~al.(1978)Black, Schweitzer, and Mandell]{Black1978}
Harold Black, Robert~L Schweitzer, and Lewis Mandell.
\newblock Discrimination in mortgage lending.
\newblock \emph{American Economic Review}, 68:\penalty0 186--91, 1978.

\bibitem[Bocian et~al.(2008)Bocian, Ernst, and Li]{Bocian2008}
Debbie~Gruenstein Bocian, Keith~S. Ernst, and Wei Li.
\newblock Race, ethnicity and subprime home loan pricing.
\newblock \emph{Journal of Economics and Business}, 60:\penalty0 110--124, 1
  2008.
\newblock ISSN 0148-6195.
\newblock \doi{10.1016/J.JECONBUS.2007.10.001}.

\bibitem[Chen et~al.(2017)Chen, Li, and Lai]{Chen2017}
Dongyu Chen, Xiaolin Li, and Fujun Lai.
\newblock Gender discrimination in online peer-to-peer credit lending: evidence
  from a lending platform in china.
\newblock \emph{Electronic Commerce Research}, 17:\penalty0 553--583, 12 2017.
\newblock ISSN 15729362.
\newblock \doi{10.1007/S10660-016-9247-2/TABLES/7}.
\newblock URL
  \url{https://link.springer.com/article/10.1007/s10660-016-9247-2}.

\bibitem[Chen et~al.(2020)Chen, Gu, Liu, and Tse]{Chen2020}
Shiyi Chen, Yan Gu, Qingfu Liu, and Yiuman Tse.
\newblock How do lenders evaluate borrowers in peer-to-peer lending in china?
\newblock \emph{International Review of Economics \& Finance}, 69:\penalty0
  651--662, 9 2020.
\newblock ISSN 1059-0560.
\newblock \doi{10.1016/J.IREF.2020.06.038}.
\newblock URL
  \url{https://linkinghub.elsevier.com/retrieve/pii/S1059056020301519}.

\bibitem[Cheng et~al.(2015)Cheng, Lin, and Liu]{Cheng2015}
Ping Cheng, Zhenguo Lin, and Yingchun Liu.
\newblock Racial discrepancy in mortgage interest rates.
\newblock \emph{Journal of Real Estate Finance and Economics}, 51:\penalty0
  101--120, 7 2015.
\newblock ISSN 1573045X.
\newblock \doi{10.1007/S11146-014-9473-0/TABLES/9}.
\newblock URL
  \url{https://link.springer.com/article/10.1007/s11146-014-9473-0}.

\bibitem[Courchane(2007)]{Courchane2007}
Marsha~J. Courchane.
\newblock The pricing of home mortgage loans to minority borrowers: How much of
  the apr differential can we explain?
\newblock \emph{Journal of Real Estate Research}, 29:\penalty0 399--439, 10
  2007.
\newblock ISSN 08965803.
\newblock \doi{10.1080/10835547.2007.12091208}.
\newblock URL
  \url{https://www.tandfonline.com/doi/abs/10.1080/10835547.2007.12091208}.

\bibitem[Munnell et~al.(1996)Munnell, Tootell, Browne, and
  McEneaney]{Munnell1996}
Alicia~H Munnell, Geoffrey M~B Tootell, Lynn~E Browne, and James McEneaney.
\newblock Mortgage lending in boston: Interpreting hmda data.
\newblock \emph{The American Economic Review}, pages 25--53, 1996.
\newblock ISSN 0002-8282.

\bibitem[Pope and Sydnor(2011)]{Pope2011}
Devin~G Pope and Justin~R Sydnor.
\newblock What’s in a picture? evidence of discrimination from prosper. com.
\newblock \emph{Journal of Human resources}, 46:\penalty0 53--92, 2011.
\newblock ISSN 0022-166X.

\bibitem[Jansen et~al.(2023)Jansen, Nguyen, and Shams]{Jansen2023}
Mark Jansen, Hieu Nguyen, and Amin Shams.
\newblock Rise of the machines: The impact of automated underwriting.
\newblock \emph{SSRN Electronic Journal}, 1 2023.
\newblock \doi{10.2139/SSRN.3664708}.
\newblock URL \url{https://papers.ssrn.com/abstract=3664708}.

\bibitem[Fuster et~al.(2022)Fuster, Goldsmith-Pinkham, Ramadorai, and
  Walther]{Fuster2022}
Andreas Fuster, Paul Goldsmith-Pinkham, Tarun Ramadorai, and Ansgar Walther.
\newblock Predictably unequal? the effects of machine learning on credit
  markets.
\newblock \emph{Journal of Finance}, 77:\penalty0 5--47, 2 2022.
\newblock ISSN 15406261.
\newblock \doi{10.1111/JOFI.13090}.

\bibitem[Hurlin et~al.(2022)Hurlin, Pérignon, and Saurin]{Hurlin2022}
Christophe Hurlin, Christophe Pérignon, and Sébastien Saurin.
\newblock The fairness of credit scoring models.
\newblock \emph{SSRN Electronic Journal}, 5 2022.
\newblock \doi{10.2139/ssrn.3785882}.
\newblock URL \url{https://arxiv.org/abs/2205.10200v1}.

\bibitem[Berk et~al.(2017)Berk, Heidari, Jabbari, Kearns, and Roth]{Berk2017}
Richard Berk, Hoda Heidari, Shahin Jabbari, Michael Kearns, and Aaron Roth.
\newblock Fairness in criminal justice risk assessments: The state of the art.
\newblock \emph{Sociological Methods and Research}, 50:\penalty0 3--44, 3 2017.
\newblock ISSN 15528294.
\newblock \doi{10.1177/0049124118782533}.

\bibitem[Calders and Verwer(2010)]{Calders2010}
Toon Calders and Sicco Verwer.
\newblock Three naive bayes approaches for discrimination-free classification.
\newblock \emph{Data Mining and Knowledge Discovery}, 21:\penalty0 277--292, 9
  2010.
\newblock ISSN 13845810.
\newblock \doi{10.1007/S10618-010-0190-X/METRICS}.
\newblock URL
  \url{https://link.springer.com/article/10.1007/s10618-010-0190-x}.

\bibitem[Dwork et~al.(2012)Dwork, Hardt, Pitassi, Reingold, and
  Zemel]{Dwork2012}
Cynthia Dwork, Moritz Hardt, Toniann Pitassi, Omer Reingold, and Richard Zemel.
\newblock Fairness through awareness.
\newblock pages 214--226. ACM Press, 2012.

\bibitem[Feldman et~al.(2015)Feldman, Friedler, Moeller, Scheidegger, and
  Venkatasubramanian]{Feldman2015}
Michael Feldman, Sorelle~A. Friedler, John Moeller, Carlos Scheidegger, and
  Suresh Venkatasubramanian.
\newblock Certifying and removing disparate impact.
\newblock volume 2015-Augus, pages 259--268, 2015.

\bibitem[Hajian and Domingo-Ferrer(2012)]{Hajian2012}
Sara Hajian and Josep Domingo-Ferrer.
\newblock A methodology for direct and indirect discrimination prevention in
  data mining.
\newblock \emph{IEEE transactions on knowledge and data engineering},
  25:\penalty0 1445--1459, 2012.
\newblock ISSN 1041-4347.

\bibitem[Kamiran and Calders(2012)]{Kamiran2012}
Faisal Kamiran and Toon Calders.
\newblock Data preprocessing techniques for classification without
  discrimination.
\newblock \emph{Knowledge and Information Systems}, 33:\penalty0 1--33, 12
  2012.
\newblock ISSN 02193116.
\newblock \doi{10.1007/S10115-011-0463-8/METRICS}.
\newblock URL
  \url{https://link.springer.com/article/10.1007/s10115-011-0463-8}.

\bibitem[Friedler et~al.(2021)Friedler, Scheidegger, and
  Venkatasubramanian]{Friedler2021}
Sorelle~A. Friedler, Carlos Scheidegger, and Suresh Venkatasubramanian.
\newblock The (im)possibility of fairness.
\newblock \emph{Communications of the ACM}, 64:\penalty0 136--143, 4 2021.
\newblock ISSN 15577317.
\newblock \doi{10.1145/3433949}.
\newblock URL
  \url{https://cacm.acm.org/magazines/2021/4/251365-the-impossibility-of-fairness/abstract}.

\bibitem[Kilbertus et~al.(2017)Kilbertus, Rojas-Carulla, Parascandolo, Hardt,
  Janzing, and Schölkopf]{Kilbertus2017}
Niki Kilbertus, Mateo Rojas-Carulla, Giambattista Parascandolo, Moritz Hardt,
  Dominik Janzing, and Bernhard Schölkopf.
\newblock Avoiding discrimination through causal reasoning.
\newblock \emph{Advances in Neural Information Processing Systems},
  2017-December:\penalty0 657--667, 6 2017.
\newblock ISSN 10495258.
\newblock URL \url{https://arxiv.org/abs/1706.02744v2}.

\bibitem[Kleinberg et~al.(2016)Kleinberg, Mullainathan, and
  Raghavan]{Kleinberg2016}
Jon Kleinberg, Sendhil Mullainathan, and Manish Raghavan.
\newblock Inherent trade-offs in the fair determination of risk scores.
\newblock 2016.

\bibitem[Narayanan(2018)]{Narayanan2018}
Arvind Narayanan.
\newblock Tutorial: 21 fairness definition and their politics.
\newblock 2018.

\bibitem[Majumder et~al.(2021)Majumder, Chakraborty, Bai, Stolee, and
  Menzies]{Majumder2021}
Suvodeep Majumder, Joymallya Chakraborty, Gina~R. Bai, Kathryn~T. Stolee, and
  Tim Menzies.
\newblock Fair enough: Searching for sufficient measures of fairness.
\newblock \emph{ACM Transactions on Software Engineering and Methodology}, 10
  2021.
\newblock ISSN 1049-331X.
\newblock \doi{10.1145/3585006}.
\newblock URL \url{https://arxiv.org/abs/2110.13029v2}.

\bibitem[Verma and Rubin(2018)]{Verma2018}
Sahil Verma and Julia Rubin.
\newblock Fairness definitions explained.
\newblock pages 1--7. IEEE Computer Society, 5 2018.

\bibitem[Ferrer et~al.(2021)Ferrer, Nuenen, Such, Cote, and Criado]{Ferrer2021}
Xavier Ferrer, Tom~Van Nuenen, Jose~M. Such, Mark Cote, and Natalia Criado.
\newblock Bias and discrimination in ai: A cross-disciplinary perspective.
\newblock \emph{IEEE Technology and Society Magazine}, 40:\penalty0 72--80, 6
  2021.
\newblock ISSN 1937416X.
\newblock \doi{10.1109/MTS.2021.3056293}.

\bibitem[Lepri et~al.(2018)Lepri, Oliver, Letouzé, Pentland, and
  Vinck]{Lepri2018}
Bruno Lepri, Nuria Oliver, Emmanuel Letouzé, Alex Pentland, and Patrick Vinck.
\newblock Fair, transparent, and accountable algorithmic decision-making
  processes: The premise, the proposed solutions, and the open challenges.
\newblock \emph{Philosophy and Technology}, 31:\penalty0 611--627, 12 2018.
\newblock ISSN 22105441.
\newblock \doi{10.1007/S13347-017-0279-X/METRICS}.
\newblock URL
  \url{https://link.springer.com/article/10.1007/s13347-017-0279-x}.

\bibitem[Wong(2020)]{Wong2020}
Pak~Hang Wong.
\newblock Democratizing algorithmic fairness.
\newblock \emph{Philosophy and Technology}, 33:\penalty0 225--244, 6 2020.
\newblock ISSN 22105441.
\newblock \doi{10.1007/S13347-019-00355-W/METRICS}.
\newblock URL
  \url{https://link.springer.com/article/10.1007/s13347-019-00355-w}.

\bibitem[Bellamy et~al.(2019)Bellamy, Mojsilovic, Nagar, Ramamurthy, Richards,
  Saha, Sattigeri, Singh, Varshney, Zhang, Dey, Hind, Hoffman, Houde, Kannan,
  Lohia, Martino, and Mehta]{Bellamy2019}
R.~K.E. Bellamy, A.~Mojsilovic, S.~Nagar, K.~Natesan Ramamurthy, J.~Richards,
  D.~Saha, P.~Sattigeri, M.~Singh, K.~R. Varshney, Y.~Zhang, K.~Dey, M.~Hind,
  S.~C. Hoffman, S.~Houde, K.~Kannan, P.~Lohia, J.~Martino, and S.~Mehta.
\newblock Ai fairness 360: An extensible toolkit for detecting and mitigating
  algorithmic bias.
\newblock \emph{IBM Journal of Research and Development}, 63, 7 2019.
\newblock ISSN 21518556.
\newblock \doi{10.1147/JRD.2019.2942287}.

\bibitem[Crenshaw(1989)]{Crenshaw1989}
Kimberle Crenshaw.
\newblock Demarginalizing the intersection of race and sex: A black feminist
  critique of antidiscrimination doctrine, feminist theory and antiracist
  politics.
\newblock \emph{University of Chicago Legal Forum}, 1989, 1989.

\bibitem[Gutterman(2022)]{Gutterman2022}
Alan~S. Gutterman.
\newblock Ageism and intersectionality: Older persons as members of other
  vulnerable groups.
\newblock \emph{SSRN Electronic Journal}, 2 2022.
\newblock \doi{10.2139/SSRN.3972842}.
\newblock URL \url{https://papers.ssrn.com/abstract=3972842}.

\bibitem[Hoffmann(2019)]{Hoffmann2019fairnessfails}
Anna~Lauren Hoffmann.
\newblock Where fairness fails: data, algorithms, and the limits of
  antidiscrimination discourse.
\newblock \emph{Information, Communication \& Society}, 22:\penalty0 900--915,
  6 2019.
\newblock ISSN 14684462.
\newblock \doi{10.1080/1369118X.2019.1573912}.
\newblock URL
  \url{https://www.tandfonline.com/doi/abs/10.1080/1369118X.2019.1573912}.

\bibitem[Atewologun(2018)]{Atewologun2018}
Doyin Atewologun.
\newblock Intersectionality theory and practice.
\newblock \emph{Oxford Research Encyclopedia of Business and Management}, 8
  2018.
\newblock \doi{10.1093/ACREFORE/9780190224851.013.48}.
\newblock URL \url{https://}.

\bibitem[Kearns et~al.(2018)Kearns, Neel, Roth, and Wu]{Kearns2018}
Michael Kearns, Seth Neel, Aaron Roth, and Zhiwei~Steven Wu.
\newblock Preventing fairness gerrymandering: Auditing and learning for
  subgroup fairness.
\newblock \emph{Proceedings of the 35th International Conference on Machine
  Learning}, 80:\penalty0 2564--2572, 7 2018.
\newblock ISSN 2640-3498.
\newblock URL \url{https://proceedings.mlr.press/v80/kearns18a.html}.

\bibitem[Das et~al.(2021)Das, Donini, Gelman, Haas, Hardt, Katzman, Kenthapadi,
  Larroy, Yilmaz, and Zafar]{Das2021}
Sanjiv Das, Michele Donini, Jason Gelman, Kevin Haas, Mila Hardt, Jared
  Katzman, Krishnaram Kenthapadi, Pedro Larroy, Pinar Yilmaz, and
  Muhammad~Bilal Zafar.
\newblock Fairness measures for machine learning in finance.
\newblock \emph{Journal of Financial Data Science}, 3:\penalty0 33--64, 9 2021.
\newblock ISSN 26403951.
\newblock \doi{10.3905/JFDS.2021.1.075}.

\bibitem[Hu et~al.(2000)Hu, Huang, Li, and Lu]{Hu2000}
Xiyang Hu, Yan Huang, Beibei Li, and Tian Lu.
\newblock Uncovering the source of evaluation bias in micro-lending.
\newblock \emph{Forty-Second International Conference on Information Systems},
  2000.

\bibitem[Stevens et~al.(2020)Stevens, Deruyck, Veldhoven, and
  Vanthienen]{Stevens2020}
Alexander Stevens, Peter Deruyck, Ziboud~Van Veldhoven, and Jan Vanthienen.
\newblock Explainability and fairness in machine learning: Improve fair
  end-to-end lending for kiva.
\newblock \emph{2020 IEEE Symposium Series on Computational Intelligence},
  pages 1241--1248, 12 2020.
\newblock \doi{10.1109/SSCI47803.2020.9308371}.

\bibitem[Kozodoi et~al.(2022)Kozodoi, Jacob, and Lessmann]{Kozodoi2022}
Nikita Kozodoi, Johannes Jacob, and Stefan Lessmann.
\newblock Fairness in credit scoring: Assessment, implementation and profit
  implications.
\newblock \emph{European Journal of Operational Research}, 297:\penalty0
  1083--1094, 3 2022.
\newblock ISSN 0377-2217.
\newblock \doi{10.1016/J.EJOR.2021.06.023}.

\bibitem[Kamiran and Calders(2009)]{Kamiran2009}
Faisal Kamiran and Toon Calders.
\newblock Classifying without discriminating.
\newblock \emph{2009 2nd International Conference on Computer, Control and
  Communication, IC4 2009}, 2009.
\newblock \doi{10.1109/IC4.2009.4909197}.

\bibitem[Teodorescu et~al.(2021)Teodorescu, Morse, Awwad, and
  Kane]{Teodorescu2021}
Mike Teodorescu, Lily Morse, Yazeed Awwad, and Gerald Kane.
\newblock Failures of fairness in automation require a deeper understanding of
  human-ml augmentation.
\newblock \emph{Management Information Systems Quarterly}, 45, 9 2021.
\newblock ISSN ISSN 0276-7783/ISSN 2162-9730.
\newblock URL \url{https://aisel.aisnet.org/misq/vol45/iss3/18}.

\bibitem[Singh et~al.(2022)Singh, Singh, Khan, and Gupta]{Singh2022}
Arashdeep Singh, Jashandeep Singh, Ariba Khan, and Amar Gupta.
\newblock Developing a novel fair-loan classifier through a multi-sensitive
  debiasing pipeline: Dualfair.
\newblock \emph{Machine Learning and Knowledge Extraction}, 4:\penalty0
  240--253, 3 2022.
\newblock ISSN 2504-4990.
\newblock \doi{10.3390/MAKE4010011}.
\newblock URL \url{https://www.mdpi.com/2504-4990/4/1/11/htm
  https://www.mdpi.com/2504-4990/4/1/11}.

\bibitem[Noble(2018)]{Noble2018}
Safiya~Umoja Noble.
\newblock \emph{Algorithms of Oppression}.
\newblock NYU Press, 2018.
\newblock URL
  \url{https://nyupress.org/9781479837243/algorithms-of-oppression/}.

\bibitem[Buolamwini(2018)]{Buolamwini2018}
Joy Buolamwini.
\newblock Gender shades: Intersectional accuracy disparities in commercial
  gender classification *.
\newblock \emph{Proceedings of Machine Learning Research}, 81:\penalty0 1--15,
  2018.

\bibitem[Stack and Meredith(2018)]{Stack2018}
Rebecca~Jayne Stack and Alex Meredith.
\newblock The impact of financial hardship on single parents: An exploration of
  the journey from social distress to seeking help.
\newblock \emph{Journal of Family and Economic Issues}, 39:\penalty0 233--242,
  6 2018.
\newblock ISSN 10580476.
\newblock \doi{10.1007/S10834-017-9551-6/TABLES/2}.
\newblock URL
  \url{https://link.springer.com/article/10.1007/s10834-017-9551-6}.

\bibitem[Andersen(2023)]{Andersen2023}
Kate Andersen.
\newblock Promoting fairness? exploring the gendered impacts of the benefit cap
  and the two-child limit.
\newblock \emph{Journal of Poverty and Social Justice}, -1:\penalty0 1--17, 3
  2023.
\newblock ISSN 1759-8273.
\newblock \doi{10.1332/175982721X16757603309669}.
\newblock URL
  \url{https://bristoluniversitypressdigital.com/view/journals/jpsj/aop/article-10.1332-175982721X16757603309669/article-10.1332-175982721X16757603309669.xml}.

\bibitem[McLanahan and Sandefur(1994)]{McLanahan1994}
Sara McLanahan and Gary~D. Sandefur.
\newblock \emph{Growing Up with a Single Parent: What Hurts, What Helps - Sara
  McLanahan, Gary D. Sandefur - Google Books}.
\newblock Harvard University Press, 1994.
\newblock URL
  \url{https://books.google.ch/books?hl=en&lr=&id=kLUX8BJ1exUC&oi=fnd&pg=PA1&dq=single+parents+discrimination&ots=Xf3TRg8dWV&sig=vLEwpw89kmnh9otWH-M7949hwKs&redir_esc=y#v=onepage&q=single%20parents%20discriminationblack&f=false}.

\bibitem[Reed(2005)]{Reed2005}
Katherine Reed.
\newblock Fairness in education for single parents in nova scotia.
\newblock 2005.
\newblock URL \url{http://www.policyalternatives.ca}.

\bibitem[Lauster and Easterbrook(2011)]{Lauster2011}
Nathanael Lauster and Adam Easterbrook.
\newblock No room for new families? a field experiment measuring rental
  discrimination against same-sex couples and single parents.
\newblock \emph{Social Problems}, 58:\penalty0 389--409, 8 2011.
\newblock ISSN 0037-7791.
\newblock \doi{10.1525/SP.2011.58.3.389}.
\newblock URL \url{https://academic.oup.com/socpro/article/58/3/389/1642853}.

\bibitem[Murchie and Pang(2018)]{Murchie2018}
Judson Murchie and Jindong Pang.
\newblock Rental housing discrimination across protected classes: Evidence from
  a randomized experiment.
\newblock \emph{Regional Science and Urban Economics}, 73:\penalty0 170--179,
  11 2018.
\newblock ISSN 0166-0462.
\newblock \doi{10.1016/J.REGSCIURBECO.2018.10.003}.

\bibitem[Dowd(1995)]{Dowd1995}
Nancy~E. Dowd.
\newblock Stigmatizing single parents.
\newblock \emph{Harvard Women's Law Journal}, 18, 1995.
\newblock URL
  \url{https://heinonline.org/HOL/Page?handle=hein.journals/hwlj18&id=25&div=&collection=}.

\bibitem[Eling et~al.(2021)Eling, Ghavibazoo, and Hanewald]{Eling2021}
Martin Eling, Omid Ghavibazoo, and Katja Hanewald.
\newblock Willingness to take financial risks and insurance holdings: A
  european survey.
\newblock \emph{Journal of Behavioral and Experimental Economics}, 95:\penalty0
  101781, 12 2021.
\newblock ISSN 2214-8043.
\newblock \doi{10.1016/J.SOCEC.2021.101781}.

\bibitem[Lundberg et~al.(1994)Lundberg, Mardberg, and
  Frankenhaeuser]{Lundberg1994}
Ulf Lundberg, Bertil Mardberg, and Marianne Frankenhaeuser.
\newblock The total workload of male and female white collar workers as related
  to age, occupational level, and number of children.
\newblock \emph{Scandinavian Journal of Psychology}, 35:\penalty0 315--327, 12
  1994.
\newblock ISSN 1467-9450.
\newblock \doi{10.1111/J.1467-9450.1994.TB00956.X}.
\newblock URL
  \url{https://onlinelibrary.wiley.com/doi/full/10.1111/j.1467-9450.1994.tb00956.x
  https://onlinelibrary.wiley.com/doi/abs/10.1111/j.1467-9450.1994.tb00956.x
  https://onlinelibrary.wiley.com/doi/10.1111/j.1467-9450.1994.tb00956.x}.

\bibitem[Worthington(2006)]{Worthington2006}
Andrew~C. Worthington.
\newblock Debt as a source of financial stress in australian households.
\newblock \emph{International Journal of Consumer Studies}, 30:\penalty0 2--15,
  1 2006.

\bibitem[Vigdor(2019)]{Vigdor2019}
Neil Vigdor.
\newblock Apple card investigated after gender discrimination complaints.
\newblock \emph{The New York Times}, 11 2019.
\newblock URL
  \url{https://www.nytimes.com/2019/11/10/business/Apple-credit-card-investigation.html}.

\bibitem[Lundberg et~al.(2017)Lundberg, Allen, and Lee]{Lundberg2017}
Scott~M Lundberg, Paul~G Allen, and Su-In Lee.
\newblock A unified approach to interpreting model predictions.
\newblock \emph{Advances in Neural Information Processing Systems}, 30, 2017.
\newblock URL \url{https://github.com/slundberg/shap}.

\bibitem[Mehrabi et~al.(2021)Mehrabi, Morstatter, Saxena, Lerman, and
  Galstyan]{Mehrabi2021}
Ninareh Mehrabi, Fred Morstatter, Nripsuta Saxena, Kristina Lerman, and Aram
  Galstyan.
\newblock A survey on bias and fairness in machine learning.
\newblock \emph{ACM Computing Surveys (CSUR)}, 54, 7 2021.
\newblock ISSN 15577341.
\newblock \doi{10.1145/3457607}.
\newblock URL \url{https://dl.acm.org/doi/10.1145/3457607}.

\bibitem[Dwork et~al.(2011)Dwork, Hardt, Pitassi, Reingold, and
  Zemel]{Dwork2011}
Cynthia Dwork, Moritz Hardt, Toniann Pitassi, Omer Reingold, and Richard Zemel.
\newblock Fairness through awareness.
\newblock \emph{Innovations in Theoretical Computer Science Conference}, pages
  214--226, 4 2011.
\newblock \doi{10.1145/2090236.2090255}.

\bibitem[Räz(2021)]{raz2021}
Tim Räz.
\newblock Group fairness: Independence revisited.
\newblock \emph{FAccT 2021 - Proceedings of the 2021 ACM Conference on
  Fairness, Accountability, and Transparency}, pages 129--137, 1 2021.
\newblock \doi{10.1145/3442188.3445876}.
\newblock URL \url{http://arxiv.org/abs/2101.02968
  http://dx.doi.org/10.1145/3442188.3445876}.

\bibitem[Hardt et~al.(2016)Hardt, Price, Price, and Srebro]{Hardt2016}
Moritz Hardt, Eric Price, Eric Price, and Nati Srebro.
\newblock Equality of opportunity in supervised learning.
\newblock \emph{Advances in Neural Information Processing Systems}, 29, 2016.

\bibitem[Hellman(2020)]{Hellman2020}
Deborah Hellman.
\newblock Measuring algorithmic fairness.
\newblock \emph{Virginia Law Review}, 106:\penalty0 811--866, 2020.
\newblock \doi{10.3886/ICPSR25103.V1}.

\bibitem[Kusner et~al.(2017)Kusner, Loftus, Russell, and Silva]{Kusner2017}
Matt Kusner, Joshua Loftus, Chris Russell, and Ricardo Silva.
\newblock Counterfactual fairness.
\newblock \emph{Proceedings of the 31st International Conference on Neural
  Information Processing Systems}, pages 4069--4079, 2017.
\newblock \doi{10.5555/3294996.3295162}.
\newblock URL \url{https://dl.acm.org/doi/10.5555/3294996.3295162}.

\bibitem[Corbett-Davies et~al.(2018)Corbett-Davies, Goel, Chohlas-Wood,
  Chouldechova, Feller, Huq, Hardt, Ho, Mitchell, Overgoor, Pierson, and
  Shroff]{corbettdavies2018}
Sam Corbett-Davies, Sharad Goel, Alex Chohlas-Wood, Alexandra Chouldechova, Avi
  Feller, Aziz Huq, Moritz Hardt, Daniel~E Ho, Shira Mitchell, Jan Overgoor,
  Emma Pierson, and Ravi Shroff.
\newblock The measure and mismeasure of fairness: A critical review of fair
  machine learning.
\newblock 7 2018.

\bibitem[Crook and Banasik(2004)]{Crook2004}
Jonathan Crook and John Banasik.
\newblock Does reject inference really improve the performance of application
  scoring models?
\newblock \emph{Journal of Banking \& Finance}, 28:\penalty0 857--874, 4 2004.
\newblock ISSN 0378-4266.
\newblock \doi{10.1016/J.JBANKFIN.2003.10.010}.

\bibitem[{IBM}()]{ibmmetrics}
{IBM}.
\newblock aif360.metrics.
\newblock URL
  \url{https://aif360.readthedocs.io/en/latest/modules/generated/aif360.metrics.ClassificationMetric.html#aif360.metrics.ClassificationMetric}.

\bibitem[Djeundje et~al.(2021)Djeundje, Crook, Calabrese, and
  Hamid]{Djeundje2021}
Viani~B Djeundje, Jonathan Crook, Raffaella Calabrese, and Mona Hamid.
\newblock Enhancing credit scoring with alternative data.
\newblock \emph{Expert Systems with Applications}, 163:\penalty0 113766, 2021.

\bibitem[Mester(1997)]{Mester1997}
Loretta~J Mester.
\newblock What's the point of credit scoring?
\newblock \emph{Federal Reserve Bank of Philadelphia: Business Review}, pages
  3--16, 1997.

\bibitem[Kim(2020)]{Kim2020}
Dongwoo Kim.
\newblock Sexism and ageism in a p2p lending market: Evidence from korea.
\newblock \emph{The Journal of Asian Finance, Economics and Business},
  7:\penalty0 537--550, 2020.
\newblock ISSN 2288-4637.
\newblock \doi{10.13106/JAFEB.2020.VOL7.NO6.537}.
\newblock URL \url{http://Creativecommons.org/licenses/by-nc/4.0/}.

\bibitem[Lee et~al.(2014)Lee, Lou, Chen, Chen, Lin, Chiang, and Chen]{Lee2014}
Eric~L. Lee, Jing~Kai Lou, Wei~Ming Chen, Yen~Chi Chen, Shou~De Lin, Yen~Sheng
  Chiang, and Kuan~Ta Chen.
\newblock Fairness-aware loan recommendation for microfinance services.
\newblock \emph{ACM International Conference Proceeding Series}, August, 8
  2014.
\newblock \doi{10.1145/2639968.2640064}.

\bibitem[Hermes and Hudon(2018)]{Hermes2018}
Niels Hermes and Marek Hudon.
\newblock Determinants of the performance of microfinance institutions: A
  systematic review.
\newblock \emph{Journal of Economic Surveys}, 32:\penalty0 1483--1513, 12 2018.
\newblock ISSN 14676419.
\newblock \doi{10.1111/JOES.12290}.
\newblock URL
  \url{https://ideas.repec.org/a/bla/jecsur/v32y2018i5p1483-1513.html
  https://ideas.repec.org//a/bla/jecsur/v32y2018i5p1483-1513.html}.

\bibitem[Maggio et~al.(2022)Maggio, Ratnadiwakara, and
  Carmichael]{dimaggio2022}
Marco~Di Maggio, Dimuthu Ratnadiwakara, and Don Carmichael.
\newblock Invisible primes: Fintech lending with alternative data.
\newblock \emph{SSRN Electronic Journal}, 5 2022.

\bibitem[Lu et~al.(2019)Lu, Zhang, and Li]{Lu2019}
Tian Lu, Yingjie Zhang, and Beibei Li.
\newblock The value of alternative data in credit risk prediction: Evidence
  from a large field experiment.
\newblock \emph{ICIS 2019 Proceedings}, 2019.

\bibitem[Óskarsdóttir et~al.(2019)Óskarsdóttir, Bravo, Sarraute,
  Vanthienen, and Baesens]{oskardottir2019}
María Óskarsdóttir, Cristián Bravo, Carlos Sarraute, Jan Vanthienen, and
  Bart Baesens.
\newblock The value of big data for credit scoring: Enhancing financial
  inclusion using mobile phone data and social network analytics.
\newblock \emph{Applied Soft Computing Journal}, 74:\penalty0 26--39, 11 2019.

\bibitem[Kim et~al.(2023)Kim, Andreeva, and Rovatsos]{Kim2023}
Savina~Dine Kim, Galina Andreeva, and Michael Rovatsos.
\newblock The double-edged sword of big data and information technology for the
  disadvantaged: A cautionary tale from open banking.
\newblock 7 2023.
\newblock URL \url{https://arxiv.org/abs/2307.13408v1}.

\bibitem[Morris and Coley(2004)]{Morris2004}
Jodi~Eileen Morris and Rebekah~Levine Coley.
\newblock Maternal, family, and work correlates of role strain in low-income
  mothers.
\newblock \emph{Journal of Family Psychology}, 18:\penalty0 424--432, 9 2004.
\newblock ISSN 08933200.
\newblock \doi{10.1037/0893-3200.18.3.424}.

\bibitem[Scott et~al.(2005)Scott, London, and Hurst]{Scott2005}
Ellen~K. Scott, Andrew~S. London, and Allison Hurst.
\newblock Instability in patchworks of child care when moving from welfare to
  work.
\newblock \emph{Journal of Marriage and Family}, 67:\penalty0 370--386, 5 2005.
\newblock ISSN 1741-3737.
\newblock \doi{10.1111/J.0022-2445.2005.00122.X}.
\newblock URL
  \url{https://onlinelibrary.wiley.com/doi/full/10.1111/j.0022-2445.2005.00122.x
  https://onlinelibrary.wiley.com/doi/abs/10.1111/j.0022-2445.2005.00122.x
  https://onlinelibrary.wiley.com/doi/10.1111/j.0022-2445.2005.00122.x}.

\bibitem[Bailey et~al.(2017)Bailey, Krieger, Agénor, Graves, Linos, and
  Bassett]{Bailey2017}
Zinzi~D. Bailey, Nancy Krieger, Madina Agénor, Jasmine Graves, Natalia Linos,
  and Mary~T. Bassett.
\newblock Structural racism and health inequities in the usa: evidence and
  interventions.
\newblock \emph{The Lancet}, 389:\penalty0 1453--1463, 4 2017.
\newblock ISSN 1474547X.
\newblock \doi{10.1016/S0140-6736(17)30569-X}.
\newblock URL \url{http://www.thelancet.com/article/S014067361730569X/fulltext
  http://www.thelancet.com/article/S014067361730569X/abstract
  https://www.thelancet.com/journals/lancet/article/PIIS0140-6736(17)30569-X/abstract}.

\bibitem[Bell(2000)]{Bell2000}
Derrick Bell.
\newblock Racism: A major source of property and wealth inequality in america.
\newblock \emph{Indiana Law Review}, 34, 2000.

\bibitem[Zambrana(2017)]{Zambrana2017}
Ruth~Enid Zambrana.
\newblock Income and wealth gaps, inequitable public policies, and the
  tentacles of racism.
\newblock \emph{American Journal of Public Health}, 107:\penalty0 1531, 10
  2017.
\newblock ISSN 15410048.
\newblock \doi{10.2105/AJPH.2017.304026}.
\newblock URL \url{/pmc/articles/PMC5607700/
  https://www.ncbi.nlm.nih.gov/pmc/articles/PMC5607700/}.

\bibitem[{UCI}()]{UCI}
{UCI}.
\newblock Uci machine learning repository: Statlog (german credit data) data
  set.
\newblock URL
  \url{https://archive.ics.uci.edu/ml/datasets/statlog+(german+credit+data)}.

\bibitem[Custers et~al.(2013)Custers, van~der Hof, Schermer, Appleby-Arnold,
  and Brockdorff]{Custers2013}
Bart Custers, Simone van~der Hof, Bart Schermer, Sandra Appleby-Arnold, and
  Noellie Brockdorff.
\newblock Informed consent in social media use – the gap between user
  expectations and eu personal data protection law.
\newblock \emph{SCRIPTed}, 10:\penalty0 435--457, 12 2013.
\newblock \doi{10.2966/SCRIP.100413.435}.

\end{thebibliography}

\newpage
\appendix{\Large \textbf{Appendix}}
\setcounter{figure}{0}
\setcounter{table}{0}
\renewcommand\thefigure{A.\arabic{figure}}
\renewcommand\thetable{A.\arabic{table}}

\begin{table}[h]
    \includegraphics[width=15cm]{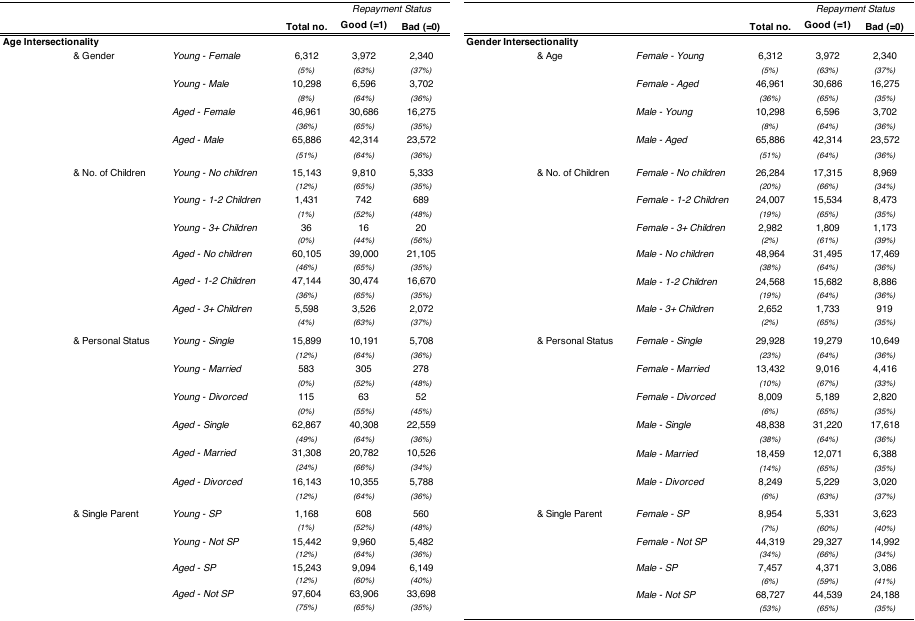}
    \centering
    \caption{Description of borrowers’ demographic characteristics and repayment status with intersectionality (Part 1).}
    \label{table:apx1}
\end{table}

\begin{table}[h]
    \includegraphics[width=15cm]{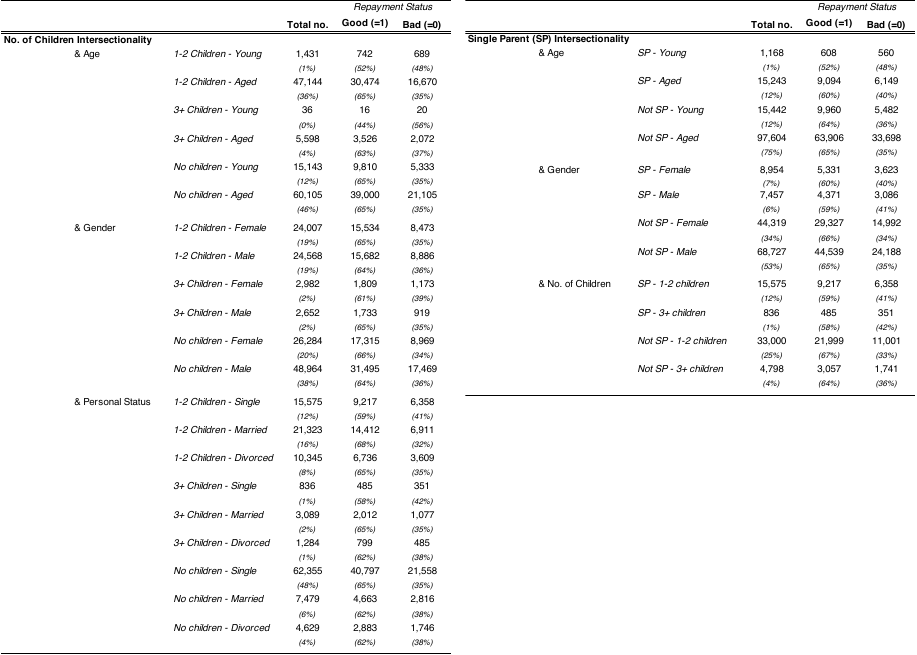}
    \centering
    \caption{Description of borrowers’ demographic characteristics and repayment status with intersectionality (Part 2).}
    \label{table:apx2}
\end{table}

\begin{table}[h]
    \includegraphics[width=7cm]{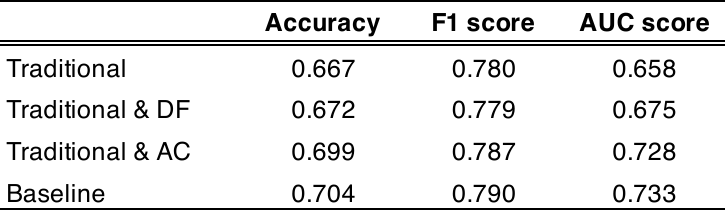}
    \centering
    \caption{Prediction model performance results.}
    \label{table:apx3}
\end{table}

\end{document}